\documentclass[12pt]{article}
\usepackage{lmodern}

\usepackage{xr}
\usepackage[margin=1in]{geometry}
\usepackage[utf8]{inputenc}

\usepackage{graphicx} 
\usepackage[square,numbers]{natbib}
\usepackage{authblk}
\usepackage{amsmath} 
\usepackage{newtxtext} 
\usepackage{hyperref}

\usepackage[table,x11names]{xcolor}
\usepackage{authblk}
\usepackage{booktabs}
\usepackage{threeparttable}
\usepackage{siunitx}
\usepackage{dcolumn}
\usepackage{rotating}

\usepackage{caption}
\usepackage{setspace}
\usepackage[font=small,labelfont=bf]{caption}

\title{Geometrics of the Adjacent Possible: \\Harvesting Values at the Curvature}
\author[1,2]{Seolmin Yang}
\author[1,2,3,*]{Hyejin Youn}
\affil[1]{Kellogg School of Management, Northwestern University, Evanston, IL}
\affil[2]{Northwestern Institute on Complex Systems, Evanston, IL}
\affil[3]{Santa Fe Institute, 1399 Hyde Park Road, Santa Fe, New Mexico 87501, USA}
\affil[*]{Correspondence can be sent to hyejin.youn@kellogg.northwestern.edu.} 

\date{\today}

\begin{document}

\maketitle

\begin{abstract}
Novelty alone is not sufficient for innovation. For new ideas and products to thrive, they must find their place within the existing societal fabric, such as institutions, conventions, and infrastructures that have been built over time. Past successes create inertia, favoring conservative advances. Here, we develop a quantitative framework to map the contours of the adjacent possible in the presence of the \textit{power of typicality}. Typical assemblies, frequently combined building blocks in past innovations, compress and curve the space of possibilities toward what is imaginable, accessible, and implementable, much like gravitational forces on new ideas and actions. We demonstrate that these curvatures in the space of possibilities are not just abstract constructs but empirically measurable through two complementary studies. We first show that Edison's inventions are primarily located in areas of high curvature, aligning with his strategy of building upon institutionalized domains. In contrast, Tesla's inventions are mainly found in low-curvature areas, indicating his propensity for exploring new territories and pushing innovation boundaries. Further analysis of the entire U.S. patent database reveals that innovations in high-curvature areas are more likely to yield monetary value. High-curvature areas indicate windows of opportunity through the interplay between innovation and convention, explaining why commercially successful ideas often emerge at the fringes of institutionalized domains.
\end{abstract}

\baselineskip24pt

\section*{Introduction}
Exploration presents both opportunities and risks. On the one hand, novel and untested exploration have the potential to yield substantial breakthroughs that drive innovation and generate new value \cite{Schumpeter1939, March1991, Fleming2001a}. On the other hand, it also comes with significant risks, such as implementation challenges and potential regulatory and legal hurdles \cite{Hargadon2001}. Novelty alone is indeed not sufficient for successful innovation; new ideas, products, and actions must find their place within the existing fabric of society, such as institutions, conventions, and infrastructures that have been built over time \cite{Schumpeter1939}. 

Successful innovations, therefore, do not emerge at random or in isolation; they arise from what is possible, imaginable, and implementable within the established socioeconomic fabric, encompassing individuals, industries, and markets \cite{Kauffman1993, Kauffman2000b, Kauffman2019, Hidalgo2007, Suarez2015}. 
But how is this explained? This socioeconomic fabric creates a complex landscape of future possibilities, making the success of innovations uneven across different directions \cite{Uzzi2013, Fink2019}. Some innovations are more likely to succeed than others \cite{Suarez2015, Uzzi2013}, some are prerequisites for others \cite{Hosseinioun2023}, and some open different sizes of adjacent possibilities, leading to waves of opportunities within their respective technological and conceptual domains \cite{Tria2014}. This landscape maps into a web of interdependent solutions, algorithms, tools, products, and services that mutually influence and shape one another \cite{Hosseinioun2023, Arthur2009, Hidalgo2021, Baldwin2000}.

Where do these underlying interdependent structures come from? Different disciplines offer different answers, and as is common in complex systems, there is no ultimate consensus. These structures are not solely attributed to immutable natural laws; social and economic constructs also shape them through aggregated individual actions and decisions from the past \cite{Arthur2009, Hidalgo2021, Kauffman2016, Carnabuci2015, Ahmadpoor2017, Kovacs2021, Pontikes2022}. 
Shared semantics or functions between components \cite{Hannan2007}, modular architectures that optimize transaction or communication costs \cite{Hidalgo2021, Baldwin2000, Delgado2014, Delgado2016, McNerney2011, Matouschek2022}, and targeted policies and incentives from formal authorities \cite{Jung2014} all contribute to structuring the possibility space. These precedents, once established and accumulated for whatever reason, are recorded in our minds, implemented in infrastructure, and codified into the legal framework, thereby shaping the underlying structure \cite{Colfer2016}. Through these processes, recurring assemblies generate salient clusters of configurations, becoming more interconnected and accessible within the broader technological and institutional landscape \cite{Arthur2009}.

We encounter the powerful presence of these structures in various aspects of our lives, such as phrases, clichés, and idioms in our language; standard recipes in our cuisine; common practices in our workplaces; legal precedents in judicial decisions; and cultural norms in our society. These can be seen as recurring assemblies of words, recipes, routinized tasks, legal procedures, and rituals. Moreover, these typical assemblies lead to common or predictable follow-ups. Just as "once upon a time" signals the start of a fairy tale, certain sequences of musical notes evoke familiar songs; soy sauce and sesame oil naturally bring rice to mind, while peanut butter and jelly bring toast; and familiar patterns in storytelling, like the hero's journey, are widely recognized \cite{campbell1949hero}. Integrated into the societal framework, these typical assemblies guide our expectations, behaviors, and even innovations in predictable ways. Analyzing them in a quantitative framework helps us understand how new ideas interact with established structures. Past decisions shape the current possibility space, and current decisions within this space reshape it for future choices.

In this paper, we develop a quantitative framework for the \textit{power of typicality} by mapping the contours of the adjacent possible in the presence of typical assemblies. Specifically, we theorize that components coalesce into widely accepted assemblies through their typicality, represented as compressed probability distances, thereby curving the space of possibilities \cite{Rosvall2008}. This power of typicality on novelty manifests in three primary ways. First, new combinations within these typical assemblies face less resistance from stakeholders, making them more feasible, fundable, and implementable \cite{Kovacs2021, Hsu2006, Bowker1999, Durkheim2009}. Second, this compression often continues until these assemblies become interlocked as epistemic norms, conventions, or even paradigms, serving as canons and modules that encapsulate rich information taken for granted in our knowledge repertoire \cite{Uzzi2013, Kuhn1962}. Third, it creates differential probability directions in their adjacent possible, guiding the paths of future innovation.

We explain that this power of typicality is why the probability of an innovation's success is not uniform across all directions but increases with proximity to existing norms and conventions, even without necessarily having formal status. This proximity creates pathways that guide our exploration and comprehension of something new, suggesting some innovations are more likely to succeed when connected with established components. As a result, through incentives, attention, and communication pressures \cite{Kovacs2021, Guilbeault2021}, the space of possibility is curved toward these attractors, much like gravitational forces acting on new ideas and actions. Make no mistake: the ``mass'' here is not physical but a construct. Nevertheless, this constructed mass wields power over our cognition, routines, organizations, infrastructure, and legal frameworks within their configuration space \cite{Berger1966}. Indeed, typical assemblies are more cognitively and institutionally accessible, serving as familiar and reliable foundations, contexts, or paradigms for future innovations \cite{Fleming2001a, Arthur2009, Kuhn1962, Shi2023}.

Adjacent to these clustered spaces, curvature emerges. These untapped areas, when adjacent to densely populated clusters, indicate windows of opportunity. As more typical assemblies occur, the possibility distance among their members becomes increasingly compressed, resulting in higher curvature and a steeper gradient between the inside and outside members. Just as physical mass curves space, this constructed mass curves the space of possibilities, making novelties more "within reach" of these powerful typical assemblies more feasible and survivable. When these assemblies serve as foundations with established solutions, these innovations, while new and untapped, become imaginable for inventors, comprehensible to stakeholders, and implementable within existing infrastructure and institutional environments \cite{Berger1966, Johnson2011, Kahneman2013, Sharma2023}. Consequently, new adjacent ideas and products in areas of high curvature can be easily built and successfully implemented, aligning with the principles of the recently successful assembly theory \cite{Sharma2023}.

We, therefore, predict that steeper curvature indicates better windows of opportunity by aligning innovations with the existing social fabric, thereby increasing social and institutional acceptance. This alignment makes new members more acceptable, integrative, and viable, paving the way for further successful exploration along established search paths \cite{Srivastava2018}. Furthermore, steeper curvature leads to waves of opportunity \cite{Tria2014}, 'hot-streaks' \cite{Liu2018}, and higher-order recombinations \cite{Shi2023}, continuously reshaping the topography of the possibility space into a more clustered arrangement with concentrated incentives, attention, and infrastructure \cite{Kovacs2021}.

We demonstrate that these curvatures in the space of possibilities are not just abstract constructs but empirically measurable. This empirical demonstration requires simplifying theoretical constructs to align with measurable quantities in available datasets. We choose recombination processes in U.S. patent records as measurable quantities in available datasets and their network representation as simplified constructs of the space of possibilities \cite{Fink2019, Youn2015, Kim2016, VanderWouden2023}. The great thing about this choice is that these constructs are widely accepted as valuable perspectives in innovation studies \cite{Arthur2009, Weitzman1998, Schumpeter1934} and are generalizable to other contexts \cite{yy2011}. 

\begin{figure*}[ht]
    \centering
    \includegraphics[width=\textwidth]{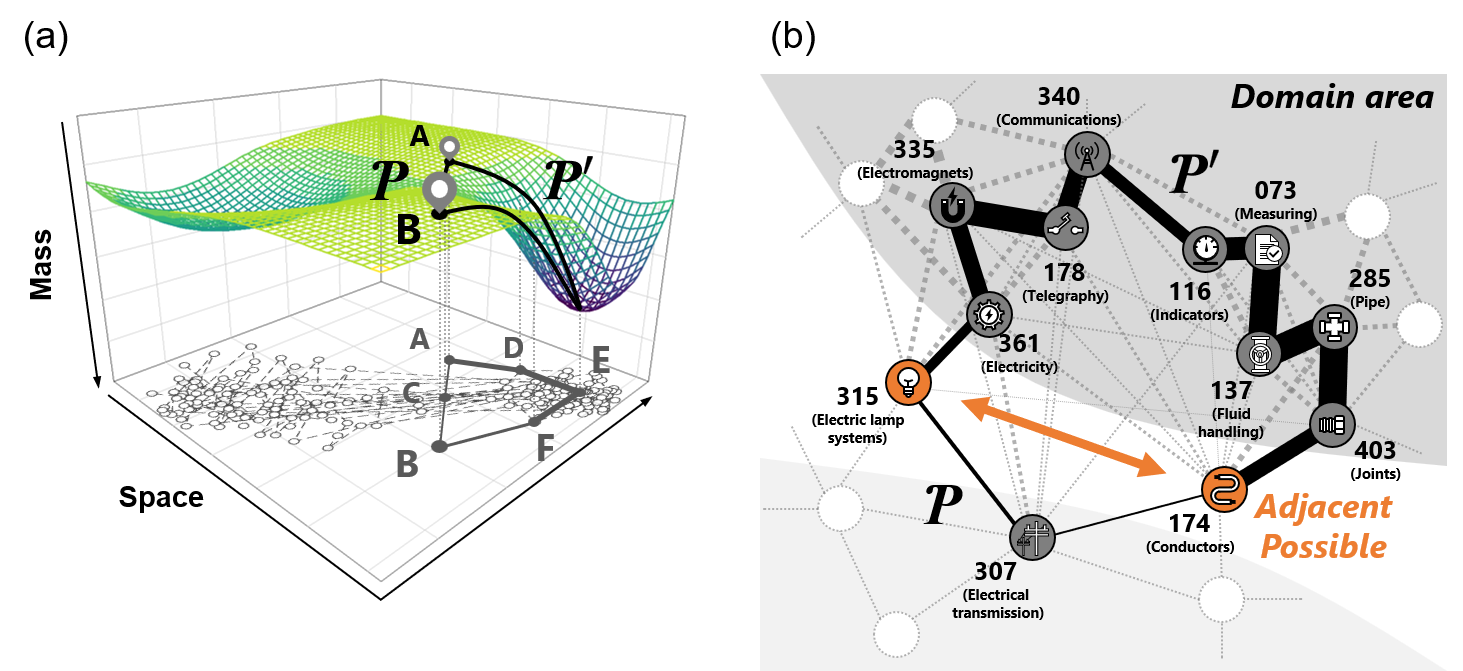}
    \caption{\textbf{Schematics of Measuring Curvature}. \textbf{(a)} The curvature of the probability space around $A$ and $B$ arises from the constructed mass of typical assemblies centered at $E$. We approximate the compression of the probability distance between two nodes by an inverse of typicality: the more typical the pair, the closer the distance. Curvature is measured by comparing the pair's least resistant path $P'$ in the presence of typicality to the expected path $P$ in its absence: $curvature = |P'|/|P|$ (see Eq. \ref{eq:eq1} and Methods). \textbf{(b)} showcases how the curvature of Edison's invention of the electric lighting system in 1881, where the `electric lamp system' (315) is firstly combined with `conductors' (174) as a new pair (marked as orange). Its adjacent spaces are compressed and resistances among corresponding components are reduced proportional to the past assemblies in patents. The more precedents, the lower the resistance among those codes. Our analysis shows the new combination's least resistant path $P'(315,361,...,403,174)$, which is five times longer degree of separations than the path without considering typical assemblies, $P(315,307,174)$. Therefore, the curvature around Edison's new pair (marked as orange) is 5.}
    \label{fig:fig1}
\end{figure*}

In this simplified model, we define the adjacent possible as a reachable space within the overall space of possibilities shaped by typical assemblies in patent records \cite{Youn2015}. Curvature is then calculated based on the effective resistance between neighboring components and the size of typical assemblies within this space \cite{Fleming2001a, Mercier2022}. Basically, the presence of more typical assemblies lowers resistance among them, creating an increasing gradient toward the outside and curving nearby probability trajectories. We approximate curvature by normalizing the number of building blocks along the least resistant paths, representing the size of easily acceptable typical assemblies (Fig. \ref{fig:fig1}). Our analysis shows that the curvature where inventions occur remains relatively stable over two hundred years of patent activities (Fig. \ref{fig:fig2}), consistent with the invariant exploration and exploitation ratio \cite{Youn2015}. However, within this variation, inventions with greater curvature have a higher chance of success (Fig. \ref{fig:fig3}) \cite{Uzzi2013}.

Indeed, the variation of curvature matters. Our findings indicate that steeper curvature provides better windows of opportunity, aligning innovations with the existing social fabric and making new ideas and products more acceptable, integrative, and viable. To test this hypothesis, we conduct two complementary studies. The first study compares the curvatures of patents by two iconic inventors, Edison and Tesla, offering initial quantitative findings and qualitative insights. While this study provides rich context, it is subject to selection bias due to its reliance on anecdotes. Therefore, the second study expands the analysis to the entire U.S. patent database, incorporating detailed information such as citations and values. This systematic approach offers more robust and generalizable findings on the relationship between curvature and invention's values. The results from the second study align with those of the first, strongly suggesting that steeper curvature leads to more useful and impactful inventions.


\section*{Results}
\subsection*{Curvature at the Adjacent Possible}
In this section, we provide an empirical approximation of the theoretical framework for the space of the adjacent possible. The first step is to simplify theoretical constructs to align with measurable quantities in the available dataset. We achieve this by using technology codes and their recombination processes in patent records as measurable quantities, with their network representation as simplified constructs of the space of possibilities \cite{Fink2019, Youn2015, Kim2016, VanderWouden2023}. Specifically, we identify assemblies as repeated bundles of technology codes (at the 3-digit level in U.S. Patent Classifiation (USPC) and the 4-level in Cooperative Patent Classification (CPC) systems) that co-appear in U.S. patent records \cite{Youn2015, Kim2016}, with ties weighted by their typicality.

While this approach may narrow our scope to a simple toy model, it remains powerful; it is both generalizable and widely accepted in innovation studies. The U.S. patent dataset is readily available and frequently used in innovation research, with recombinations central to the field \cite{Arthur2009, Weitzman1998, Schumpeter1934}. Network representation and its structural properties are well-studied for such processes \cite{Hidalgo2007, Tria2014}. Therefore, these choices serve as the typical assemblies for our study. Finally, our framework is generalizable beyond technology innovation and is designed to be applicable to innovations in cultural evolution as well.

Figure \ref{fig:fig1}a illustrates our model of how possibility distance is compressed and the geometry of the adjacent possible for a new combination is curved. We first approximate this compressed distance in probability space using network resistance \cite{Mercier2022}, suggesting that inventors and stakeholders take the path of least resistance along these nodes of typical assemblies. Specifically, the probability distance between $A$ and $B$ is approximated by the effective resistance $d_{AB} = 1/n_{AB}$, inversely proportional to their typicality $n_{AB}$, that is, the more typical combination is, the less resistance, and the shorter distance.

In fact, the effective resistance measure in network science is nearly identical to the familiarity metric used in innovation studies, with one key difference: the depreciation of past actions \cite{Fleming2001a}. Basically, it accounts that the relevance of past events to the current state of familiarity decreases exponentially over time:
\begin{equation}
d_{AB} = 1/{\sum_{\Delta_t} n_{AB}(\Delta_t) e^{-\Delta_t/t_c}}
\label{eq:eq1}
\end{equation}
where $n_{AB} (\Delta_t)$ represents the past precedents over $\Delta_t$ years, whose effects depreciate exponentially at rate $t_c$. Following the convention that the typical patent citation cycle spans approximately ten years \cite{Fleming2001a}, we take $t_c$ for five years.

Note that the newly created pair, $A$ and $B$, are not isolated in Fig. \ref{fig:fig1}a. Nearby, there is a large, dense cluster centered at $E$, where distances can be further reduced in our cognitive space, organizational structure, or physical infrastructure through their typicality, creating more compression in the adjacent space \cite{Mercier2022, Bozzo2013}. Within these clusters, one member naturally leads to another, creating predictable follow-ups, as if their possibility distance is compressed in our cognitive or invention processes. These typical assemblies near the focal pair further reduce the effective resistance for the newly created pair ($A-B$), such that $d_{AB} \sim n_{AB}^{-1} \ll \sum_{i,j\in P'} n_{ij}^{-1}$, where $P'$ passes through the members of typical assemblies, $i$ and $j$. This results in a non-trivial least resistance path $P'(A,D,E,F,B)$ that initially appears longer than $P(A,C,B)$, the least resistance path in the absence of typicality, as shown in Fig. \ref{fig:fig1}a.

At the periphery of large dense clusters, curvature emerges. In mathematics, curvature is defined as the degree to which a curve deviates from a straight line in the absence of mass. Here, only the mass is not physical but rather conceptual, \textit{constructed} from historical events and data. To measure curvature near novelty, we calculate the least-resistant path between a newly created pair, normalized by the expected path in the absence of typicality, as follows:

\begin{equation}
curvature_{A,B} = \frac{|P'_{A,B}|}{|P_{A,B}|}
\label{eq:eq2}
\end{equation}

For example, in Fig. \ref{fig:fig1}a, $curvature_{A,B}$ = 2 because $|P'_{A,B}| = 4$ through the dense clusters, $D$, $E$ and $F$, normalized by the expected distance in the absence of typicality, $|P_{A,B}|=2$. When there is no cluster and the probability of combination and feasibility is equal in any direction, the area is flat, and thus curvature is one as $P = P'$. 

\subsection*{Curvatures of Edison's and Tesla’s Adjacent Possible}
To contextualize our theoretical framework, we selected Thomas Edison as an empirical case study and compared him with Nikola Tesla using our methodology. Both are prominent figures from the same era and are frequently compared anecdotally and scholastically \cite{Hargadon2001}. Edison, often seen as a successful entrepreneur and businessman, captured significant value through his well-known design strategy. Conversely, Tesla's achievements were not sufficiently rewarded at the time for his innovative inventions. Here, we investigate whether the curvature in their invention activities reveals their systematic differences.

\begin{figure*}[ht]
    \centering
    \includegraphics[width=\textwidth]{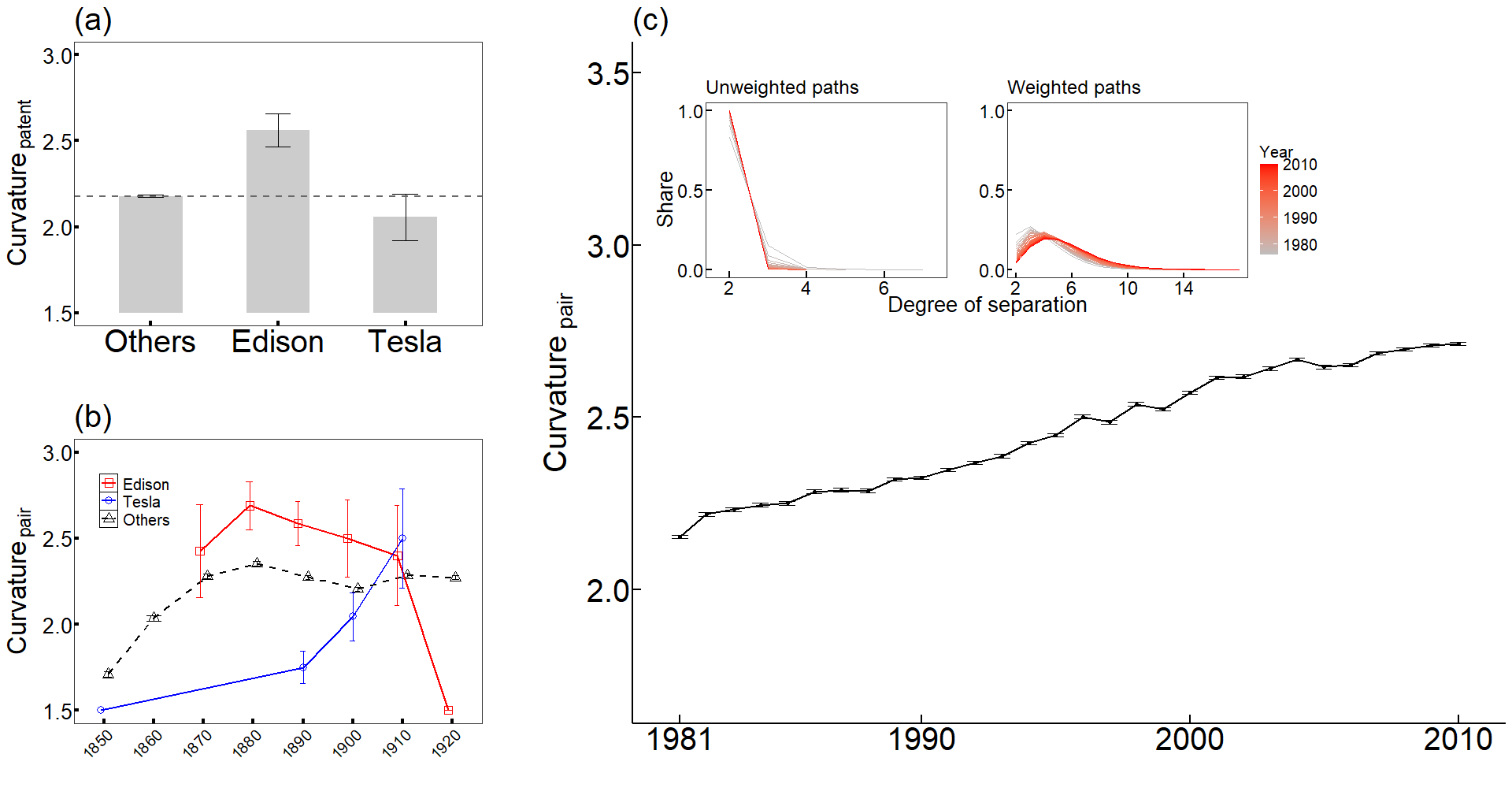}
    \caption{\textbf{Curvature around Newly Created Pairs.} \textbf{(a)} shows the average curvature in the patents of Edison and Tesla. \textbf{(b)} show the average curvature of novel pairs in U.S. patents from 1845 to 1925 (black triangle), along with the average curvature around pairs created by Edison (red square) and Tesla (blue circle) (see Methods and SI Sec-\ref{si:sec3}). \textbf{(c)} shows the average curvature with standard errors for novel pairs from 1981 to 2010. The insets show the distribution of least-resistant path lengths in the presence $P'$ and absence $P$ of typical assemblies (weighted ties by Eq. \ref{eq:eq1}}
    \label{fig:fig2}
\end{figure*}

We first examine the curvature of Edison's well-studied patent \cite{Hargadon2001} for its strategic position in the adjacent possible space. This focal patent involves the system of electric lighting, creating a new pair by combining \textit{electric lamp and discharge devices} (315) with \textit{conductors and insulators} (174) for the first time.  Figure \ref{fig:fig1}b illustrates the adjacent possible space around these technology codes (marked as orange nodes), showing other nodes directly and indirectly connected to them. Some are tightly clustered, while others are loosely connected.

These typical assemblies compress the distance around them, according to Eq. \ref{eq:eq1}, and we calculate the patent's curvature using Eq. \ref{eq:eq2} (see Methods and SI Sec-\ref{si:sec2}). Our analysis shows that the effective resistance paths between the new pair (orange) include nine intermediate nodes, i.e., $|P'|=10$, normalized by $P=2$ through 307. Thus, the focal patent's curvature is five, exceptionally steeper than most curvatures of the time (including Tesla's), as shown in Figs. \ref{fig:fig1}b.

What are these typical assemblies that have made such a steep curvature near Edison's new invention of electric lighting systems? Examining these typical assemblies up close reveals that they encompass a range of well-established technological domains. These include tightly clustered technologies related to telegraphy (codes 361, 335, 178, and 340) on the top left of the large cluster and those comprising gas infrastructures (codes 116, 073, 135, 285, and 403) on the top right, as shown in Fig. \ref{fig:fig1}b. These typical assemblies have made the new orange pair less resistant than what is expected, indicated by the exceptionally higher curvature. 

Initially, the inclusion of gas pipes and joint assemblies in the least resistant paths $P'$ for electric lighting systems might seem unrelated, raising concerns about the methodology. However, an examination of Edison's patent claims (US251551) reveals his strategic decision to leverage stakeholders' familiarity with the existing gas industry. He described how his new system would ``utilize the gas burners and chandeliers now in use. In each house, I can place a light meter, whence these wires will pass through [existing gas pipes in] the house, tapping small metal contrivances that may be placed over each burner" \cite{Hargadon2001}. Remarkably, our least resistance path method identifies these hidden institutional connections without any contextual, not even textual, analysis. We used only structures that emerged from past assemblies in patent records and did not account for any records post-invention. Contexts are indeed understood as structural properties.

What is more interesting is found in Edison's other patent for the initial introduction of the phonograph (US465972). This invention was not immediately successful, as it failed to align with practical applications and convince the public of its utility \cite{Hargadon2001, Conot1979}. Our method indicates this with a relatively lower curvature of 3.5 compared to the electric lighting systems' curvature of 5, demonstrating how the patent was embedded in the space of possibilities. This supports the notion that invention value is most likely to be created at the periphery of clusters, where high curvature exists, rather than within the cluster or far from it \cite{Carnabuci2015}.

Within the steep curvature of the electric lighting systems, smaller, interconnected clusters are observed, such as those related to telegraphy, gas infrastructure, and electric systems. These distinct yet subtly connected clusters indicate nested structures arising from the recursive nature of combinatorial innovation processes \cite{Fink2019, Arthur2009}. Nested clusters often exhibit lower curvature, as they are already embedded within the larger domain cluster, generating little variation in resistance paths and resulting in minimal curvature and nearly flat terrain.

Edison's inventions, created in high curvature spaces, align with his inventive strategy and contrast with Tesla's approach, calling for a systemic comparison between the two iconic inventors in the adjacent possible. Figures \ref{fig:fig2}a and \ref{fig:fig2}b show the average curvature around the novel combinations made in Edison's hundred patents and Tesla's seventeen patents (see Methods and SI Sec- \ref{si:sec4}). Analyzing the patents from 1836 to 1933 and their curvatures in the space of possibilities at that time, we find that Edison explored exceptionally high-curvature spaces, surpassing most inventors of his era.

\subsection*{Curvatures and Values}
A comparison of Edison's patents with Tesla's reveals their different strategies for capturing and creating value, which can be understood through the curvature of their positions in the space of possibilities. Figures \ref{fig:fig2}a and \ref{fig:fig2}b show Edison's patents are mostly embedded in high curvature areas, suggesting that he and his colleagues strategically positioned their inventions near the peripheries of well-accepted and institutionalized clusters of typical assemblies that have been implemented, registered, and integrated within existing infrastructures and conventions \cite{Hargadon2001}. Positioned near these well-established institutional clusters, Edison's inventions were thus readily imaginable, acceptable, and implementable for stakeholders of the time, enabling him to capture and create value from his innovations effectively \cite{Hargadon2001}.

In contrast, Tesla's patents are situated in lower curvature areas compared to the average of his time. This perhaps explains why Tesla received relatively fewer rewards than Edison; his inventions were not as closely aligned with the established norms and conventions of the era. Positioned further away from the clusters of typical assemblies in flatter terrain, it was more challenging for him to capture and create value from his innovations.

These historical comparisons between Edison and Tesla suggest that curvature plays a crucial role in determining an inventor's ability to create and capture value from their inventions. By strategically positioning their patents near the peripheries of well-accepted and institutionalized clusters, inventors can leverage existing infrastructures and conventions to maximize the impact and success of their innovations. The curvature of an invention's position in the space of possibilities serves as a valuable indicator of its potential to capture and create value within its historical context.

While the historical data from Edison and Tesla's eras provide contextual insights through detailed qualitative and quantitative analyses, they do not offer systematic evidence for the impact of curvature beyond anecdotal cases. Additionally, our findings might be subject to sampling bias as our insights rely on only two cases. To gain a more comprehensive understanding of how curvature in the adjacent possible space interacts with impact metrics, such as citations and market value, we evaluate these findings using a more recent dataset for US-granted patents filed from 1981 to 2010 where citations and market value estimations are available \cite{Kogan2017}.

Figure \ref{fig:fig2}c shows that patent curvatures remain relatively stable over two hundred years of patent activities, consistent with the recent finding of the invariant exploration and exploitation ratio \cite{Youn2015}. Additionally, the constructed adjacent possible maintains robust structural properties regardless of the classification system used: USPC in Fig. \ref{fig:fig2}b and CPC in Fig. \ref{fig:fig2}c. However, within this stable average, there are huge variations in curvatures, indicating that inventors in different areas may be rewarded differently, just like Edison and Tesla. 

To further understand this phenomenon, we systematically examine how inventions with greater curvature have a higher chance of success using modern datasets where detailed patent information is parsed and estimated. We estimate an invention's value by analyzing patent forward citations within five and ten years following the grant \cite{Ahuja2001, Albert1991, Trajtenberg1990}. Hit patents are identified as those whose citations fall within the top 5th percentile of their grant year cohort, and a dummy variable is assigned a value of 1 for these hit patents.

We recognize that technological value does not always translate into market value. This discrepancy can arise from a time lag between the creation of technological value and its market recognition or because a novel idea might spur future innovations without receiving immediate recognition, a phenomenon known as the "sleeping beauty" effect \cite{Ke2015sleeping}. Therefore, while citation counts are highly correlated with market value, they do not necessarily reflect it \cite{Kogan2017}. Distinguishing between novelty and innovation is crucial to ensure that novel ideas are appreciated and implemented in the real world.

We, therefore, measure value capture separately from value creation to account for potential discrepancies between them. For value capture, we use a patent's market value in U.S. dollars \cite{Kogan2017}. We then compare these market values with the invention's position in the space of possibilities to identify any correlations or patterns. This differentiation helps distinguish between Tesla's focus on creating technological value and Edison's ability to capture market value effectively. By considering both technological value (measured by citations) and market value (measured by stock market reactions) in relation to an invention's position, we gain a more nuanced understanding of the factors influencing an invention's success and impact. Importantly, these aspects are not mutually exclusive; Edison was able to both create and capture value by strategically positioning himself in high-curvature spaces.

\begin{figure*}[ht]
    \centering
    \includegraphics[width=0.9\textwidth]{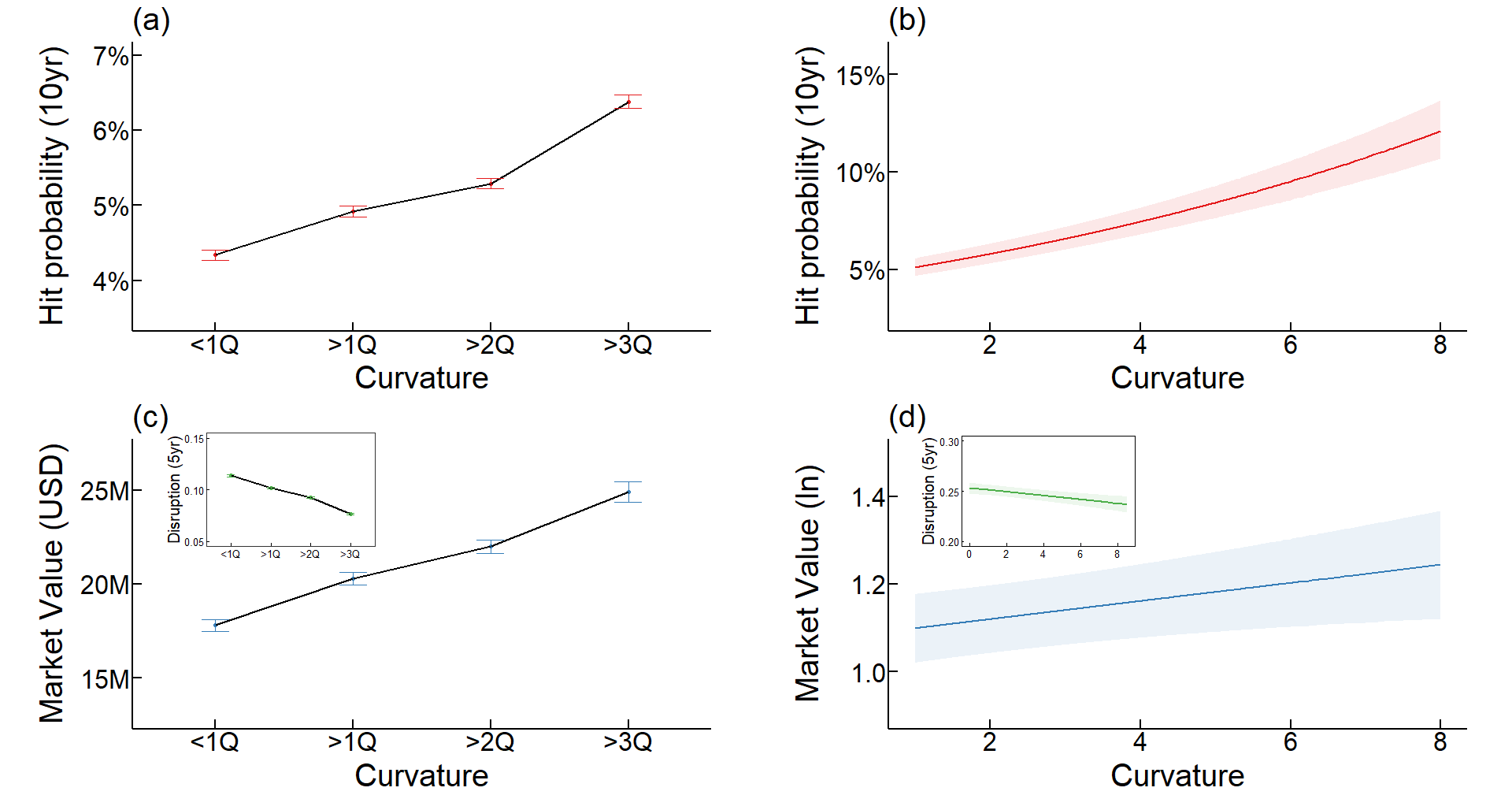}
    \caption{\textbf{Curvature and Invention Values.} \textbf{(a)} and \textbf{(b)} display the relationship between patent's curvature and hit probability, respectively, divided by quartiles and analyzed using the logit regression in Model 1 of Table \ref{tab:tab1}. 
    \textbf{(c)} and \textbf{(d)} show the relationship between patent's curvature and market value by quartiles, divided by quartiles and analyzed via the OLS regression in Model 2 of Table \ref{tab:tab1}, respectively. Hit probability refers to the likelihood of a patent's citation to be in the top 5\% of its cohort, and market value (in millions of nominal dollars) is estimated as the increase in the market value attributable to the patent \cite{Kogan2017}. The insets show the relationship between patent's curvature and their disruption index \cite{Funk2017} in the OLS regression Model 3 of Table \ref{tab:tab1}. Whiskers on the graphs denote the mean and standard error, while shaded areas represent the 95\% confidence interval }
    \label{fig:fig3}
\end{figure*}

\newpage
\begin{table}[!ht] \centering 
  \caption{Regression Results for Predicting the Invention Value} 
  \label{tab:tab1} 

\begin{threeparttable}
\begin{tabular}{@{\extracolsep{5pt}}lD{.}{.}{-3} D{.}{.}{-3} D{.}{.}{-3} } 
\\[-1.8ex]\hline 
\\[-1.8ex] & \multicolumn{1}{c}{Hit Probability (10yr)} & \multicolumn{1}{c}{Market Value (ln)} & \multicolumn{1}{c}{Disruption (5yr)} \\ 
\\[-1.8ex] & \multicolumn{1}{c}{Model 1} & \multicolumn{1}{c}{Model 2} & \multicolumn{1}{c}{Model 3}\\ 
\hline \\[-1.8ex] 
 Constant & -5.171^{***} & 0.460^{***} & 0.422^{***} \\ 
  Forward citations (10yr) (ln) &  & 0.092^{***} & 0.031^{***} \\ 
  \# patent references (ln) & 0.613^{***} & 0.462^{***} & -0.108^{***} \\ 
  Non-patent references & 0.552^{***} & 0.171^{***} & 0.012^{***} \\ 
  \# CPC main group & 0.071^{***} & 0.0005 & 0.001^{***} \\ 
  Contrast & 0.611^{***} & -0.801^{***} & 0.003 \\ 
  Primary code size (ln) & 0.105^{***} & 0.051^{***} & 0.0001 \\ 
  Novelty proportion & -0.909^{***} & 0.217^{***} & 0.021^{***} \\ 
  Familiarity (ln) & 0.049^{***} & -0.037^{***} & -0.0004 \\ 
  \textbf{Curvature} & 0.135^{***} & 0.021^{**} & -0.002^{***} \\ 
 \hline \\[-1.8ex] 
Application Year & Y & Y & Y \\ 
Primary CPC Section & Y & Y & Y \\ 
Observations & \multicolumn{1}{c}{357,964} & \multicolumn{1}{c}{100,345} & \multicolumn{1}{c}{344,849} \\
McFadden's $R^{2}$ & 0.134 &  &  \\ 
AIC & \multicolumn{1}{c}{126,063} &  &  \\  
R$^{2}$ &  & \multicolumn{1}{c}{0.103} & \multicolumn{1}{c}{0.183} \\ 
Adjusted R$^{2}$ &  & \multicolumn{1}{c}{0.103} & \multicolumn{1}{c}{0.183} \\ 
\hline 
\end{tabular} 
\begin{tablenotes}[para, flushleft] 
  $^{\dag}$\textit{p}$<$0.10; $^{*}$\textit{p}$<$0.05; $^{**}$\textit{p}$<$0.01; $^{***}$\textit{p}$<$0.001
\end{tablenotes}

\end{threeparttable}

\end{table}

Figures \ref{fig:fig3}a and Table \ref{tab:tab1} support our initial findings of the Edison and Tesla cases, showing that patents with higher curvature in their new combinations have a higher probability of success (hit probability). These results persist even when controlling for various known predictors of an invention's value, including application year, reference size \cite{Harhoff2003}, non-patent references \cite{Ahmadpoor2017, Poege2019}, the number of unique assigned codes \cite{Lerner1994}, and representative codes assigned to a given patent \cite{Kovacs2021}. Model 1 of Table \ref{tab:tab1} reveals a positive coefficient for curvature ($\beta = 0.135$, $p < 0.001$), indicating that a one-unit increase in curvature is associated with a 14\% ($e^{0.135} - 1$) increase in the odds of being a hit patent within a ten-year window (see Methods and SI Sec-\ref{si:sec4}).  
 
This positive correlation holds even with value capture as shown in Fig. \ref{fig:fig3}c and \ref{fig:fig3}d. Inventions at high curvature are associated not only with high citations but also with higher market values in nominal dollars, as shown in both quartile and regression results ($\beta = 0.021$, $p < 0.001$), indicating that a one-unit increase in the average curvature results in approximately a 2\% increase U.S. dollars (see Model 2 of Table \ref{tab:tab1}).

Our findings for value creation and value capture demonstrate robustness across various alternative measures for the dependent variables. Specifically, the results remained consistent when using different time windows for hit probability, such as five years instead of ten years. The findings held true for alternative measures of impact, including total forward citations within both five- and ten-year windows. Finally, Our market value findings remained stable when adjusted to 1982 million dollars using the Consumer Price Index (CPI) \cite{Kogan2017}. In all these variations, the results consistently supported our primary analysis conclusions (see robustness analyses in SI Sec-\ref{si:sec4}). 

While proximity to typical assemblies can increase the potential for an invention's practical usage and, thus, monetary rewards, it may not result in disruptive innovations. To explore this relationship, we examined the relationship between curvature and the disruptiveness of inventions \cite{Funk2017}.
Our analysis reveals an inverse relationship between curvature and disruptiveness. This is evidenced by the negative slope in the insets of Fig. \ref{fig:fig3} and the negative coefficient in Model 3 of Table \ref{tab:tab1} ($\beta = -0.002$, $p < 0.001$). Both indicate that inventions arising at high curvatures tend to be less disruptive than those emerging at low curvatures. 

Indeed, inventions benefit from curvature in terms of practical application and value capture, but this doesn't necessarily translate to disruption of the existing societal fabric. Our findings suggest that high curvature can facilitate the creation and capture of value, although it may not lead to highly disruptive innovations.

These findings align with established insights in several ways. Building on typical assemblies for innovation can decrease the risk of failure by avoiding high-risk and uncertain scenarios. Although less radical, inventions with these typical assemblies that form well-defined interconnections within their domains can be considered imaginable for inventors and highly profitable to the market. This factor is crucial, especially considering that inventions exploring entirely new possibilities often face technological uncertainties, adversely affecting their probability of success \cite{Utterback1971, Klepper1997, Simonton2004}.

\subsection*{Curvature, Contrast, Search, and Attention}
Some may argue that the concept of probability distance in the space of possibility using resistance is abstract and lacks direct contextual grounding in micro-socioeconomic mechanisms. To address this, we employ a well-established contrast measure \cite{Kovacs2021} to connect probability distance to search and attention mechanisms within the innovation context, aligning it more closely with social mechanisms \cite{Ocasio2011attension}.

For an invention to make an impact, it must attract attention by being easily searchable amid information overload \cite{Kovacs2021, Ocasio2011attension, Barabasi2018}. Contrast theory posits that clearly defined categories enhance an invention's searchability, leading to more frequent citations \cite{Kovacs2021}.  Kov{\'{a}}cs et al. \cite{Kovacs2021} provide micro-level observations of patent examiners' code assignment and prior art search processes, showing that higher contrast in a technology domain increases its likelihood of being used in searches, thus boosting citations. In probability terms, this indicates a curved adjacent possible, where search likelihood is directed toward well-defined categories.

We calculated the contrast following \cite{Kovacs2021} to compare it with our curvature measure. Contrast is defined as the proportion of patents that are exclusively classified by codes within an upper-level category that codes belong to. This structural property aligns with our concept of typical assemblies, which create gradient boundaries resulting in steep curvature. Using the Cooperative Patent Classification (CPC) tree, we measure contrast for coarse-grained subclasses (one level up of codes). We describe the details in Methods and SI-Sec \ref{si:sec2}. 

Figure \ref{fig:figS10} shows a positive correlation between contrast and curvature measures, explained by their structural similarity to the modularity measure in network science \cite{Newman2006modularity}. Unlike contrast, our curvature measure doesn't presuppose membership; instead, denser clusters create bottom-up membership, generating curvature. This means codes are connected within their main group levels, whether presupposed or emergent.

\begin{figure*}[ht]
    \centering
    \includegraphics[width=\textwidth]{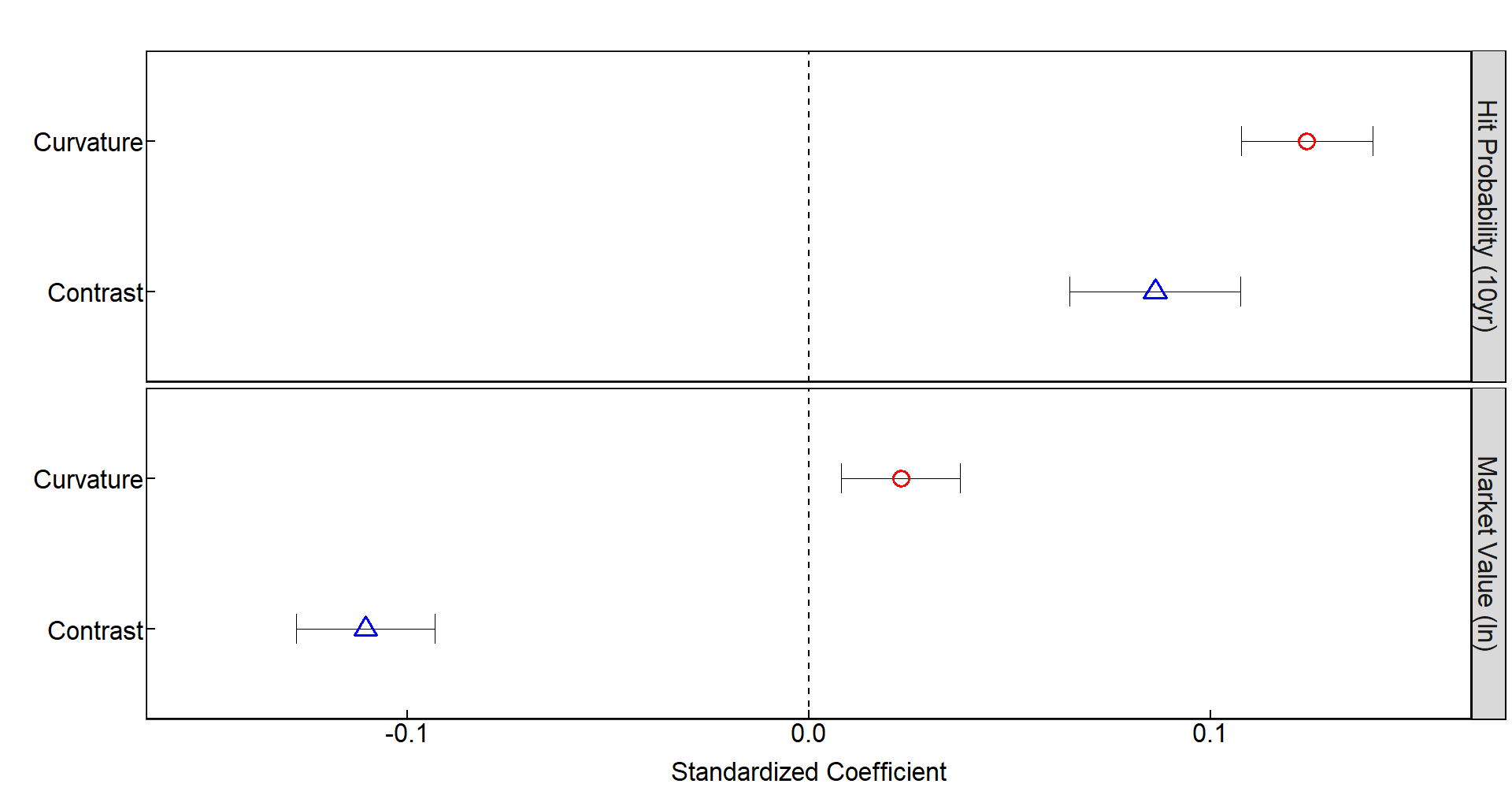}
    \caption{\textbf{Comparing Curvature and Contrast}
    The coefficients of these attributes, along with their 95\% confidence intervals, are estimated by regression models (Models 1 and 2 of Table \ref{tab:tab1}).}
    \label{fig:fig4}
\end{figure*}

Figure \ref{fig:fig4} compares standardized coefficients for contrast and curvature. Both measures are positively correlated with hit probability, as expected from their positive correlation in Fig. \ref{fig:figS10}. However, their effects go the opposite paths for value capture (market value responses): curvature increases both hit probability and market values, while contrast increases hit probability but decreases market values.

Overall, the category's contrast can be generalized into our curvature measure by not presupposing the hierarchical classification system that each code is embedded in. The hierarchical structure can be reconstructed from coarse-graining codes' connectivity. As such, our approach relaxes the top-down construct and instead examines connectivity at the bottom level. Assemblies with high curvature, characterized by a steep gradient from internal to external connections, typically result in high contrast. However, this alone is not sufficient.
 
\section*{Discussion}
In this study, we investigated the impact of novel possibilities adjacent to typical assemblies where ``reality of everyday life’’ domains were characterized as significantly curved areas \cite{Berger1966}. The space of the adjacent possible at high curvature indicates the higher-order recombinants (``long-jump’’ combinations between distant knowledge areas) allowed by compressed adjacent possible in the information infrastructure \cite{Kauffman1993, Levinthal1997, Kneeland2020}. In addition, these curvatures are the very spot aligned with collective actions, including prevailing norms and established institutions, thus representing the power of typicality. We defined curvature as the relative degree to which a curve deviates from a straight line by comparing the least-resistance paths and the expected distance in the absence of typicality. We found that novel possibilities arising from the high curvatures were put more into practical use and commercialization. Our findings suggest that positioning new ideas at high curvature is readily recognized and accepted by inventors, examiners, and the public.

By shifting our focus from novelty and originality to the power of typicality, we contribute to improving our understanding of how past innovations shape the space of their surrounding possibilities \cite{Lee2024} and how imaginable and recognizable the nearby possibilities are within existing institutions. Novel possibilities often break away from established paradigms \cite{Ferguson2017a}, showing non-paradigmatic characteristics \cite{Wang2019b}. These innovations face the challenge of gaining consensus on their value within the existing paradigmatic system. The key to harvesting value from these novel innovations lies in leveraging established practices while introducing a moderate novelty \cite{Hargadon2001}. By placing new possibilities in proximity to high curvature where technological clusters are institutionalized, we can reduce the risks associated with introducing new concepts and lower the barriers to market entry. Strategic placement at such an opportune time significantly increases the likelihood of successfully realizing value from innovations \cite{March1991, Suarez2015, Uzzi2013, Kim2016, Kuhn1963}.

Our quantitative framework has the potential to broaden methodological implications for the recent surge in machine learning. From a broad perspective, data-driven methodologies leverage context embedding through the co-occurrence of vocabularies, often referred to as a continuous bag of words \cite{Mikolov2013}. This approach allows us to adapt machine learning techniques to analyze the co-occurrences of technology codes in inventions. Fundamentally, our data input mirrors this approach as discussed in \cite{Tacchella2020}, and the emergence of the context in \cite{Shi2023}. Similar to semantic domains, commonly used and combined technologies are found in close proximity, providing a valuable set of tools for inventors. These domains indicate the unevenness of the space of the known the possible, constraining, and promoting behaviors of agents (inventors, scientists, firms, and universities). Given that embeddings are constructed to capture the meaning of singular words or phrases, our method can provide a good first approximation for embedding space that predicts prescient idea, emerging the periphery \cite{Vicinanza2022}, and the likelihood of innovative connections \cite{Tacchella2020}.

\section*{Methods}

\subsubsection*{Data and Sample}
We use two data sources for patent records from the United States Patent and Trademark Office (USPTO): the Google Patent Dataset and PatentsView. The Google Patent Dataset offers extensive historical records of assigned technological codes and their grant years from 1791 to 2015, including the first patent. However, it lacks detailed information such as application years, citations, inventor information, and market value data. Therefore, we use Google Patents' historical data to analyze Edison's and Tesla's patents and PatentsView, which includes U.S. granted patents since 1976, for more comprehensive analyses.

Specifically, we analyze 2,815,226 granted patents between 1836 and 1935, with 1084 patented by Edison and 112 by Tesla. For patents granted between 1976 and 2023, we use 4,710,933 utility patents filed between 1975 and 2010, dropping the last 13 years, to ensure sufficient time for citations to be granted. For value creation and capture, we use Kogan et al.'s dataset, which includes 100,345 patents with market values \cite{Kogan2017}, and Funk and Owen-Smith's, which includes 344,849 patents with a disruption index \cite{Funk2017, FunkData}(see SI-Sec \ref{si:sec1} for details).

\subsection*{Measures}

\textbf{Curvature:} We calculate the curvature for each newly created pair of technology codes in a patent applied in year $t$. The CPC system has five levels: section, class, subclass, main group, and sub-group. Our analysis focuses on the main group, which is the second to last level of granularity. When a patent creates more than one new pair, we average their curvatures to represent the patent's overall curvature. To measure each pair's curvature, we first calculate resistances between codes using Eq. \ref{eq:eq1}, where $d_{AB}$ denotes the number of patents combining codes $A$ and $B$, with the depreciation rate $t_c$ set to five years, following \cite{Fleming2001a}. Due to limitations in historical data, we use grant years instead of application years for analyses of Edison's and Tesla's patents.

After establishing the technology network with $d_{AB}$ in Eq. \ref{eq:eq1}, we calculate the least resistant path $P'$ between nodes combined by the focal patent, using a standard shortest path algorithm \cite{Dijkstra1959}. To ensure a fair comparison, we normalize $|P'|$, the degree of separation, with the expected distance in the absence of resistance ($|P|$). Essentially, the curvature of a patent's new pair reflects a relatively higher degree of separation when resistance is considered
(see SI-Sec \ref{si:sec2} for details).

\textbf{Invention Values:} We measure the invention’s value creation as (1) the probability of hit patent, (2) the number of forward citations, and value capture as (3) the stock market value reaction \cite{Kogan2017, Ahuja2001, Trajtenberg1990, Albert1991, Funk2017}. The hit probability is defined as being in the top 5th percentile by citations among the application year cohort. It is a dummy that took the value 1 for a hit patent. We account for citations only five- and ten-year windows after the granted year of the patent to ensure fair comparisons, but our analysis is robust to a yearly time window. For the invention's value capture, we use Kogan et al.'s KPSS index, which measures the private value of patented inventions by estimating stock market reactions to them \cite{Kogan2017}. We use logarithm-transformed value for our regression analysis. We use Funk and Owen-Smith's disruption index \cite{Funk2017} to discern the extent to which a patented invention consolidates or destabilizes knowledge flows (see SI-Sec \ref{si:sec2} for details).

\textbf{Contrast:} We calculate each code's contrast and assign this value to each patent when it is the patent's primary code \cite{Kovacs2021}. Contrast is determined by how codes are assigned within patents. When multiple codes appear in a patent, a primary class (a higher-level category) is identified. If all codes in a patent belong to the primary class, the patent's recombinativeness is one. If only one code belongs to the primary class, the recombinativeness is one divided by the total number of codes in the patent. The contrast for each patent class is calculated as one minus the average recombinativeness of the patents under this primary class in a given year. A patent's contrast is one when all patents classified under it are exclusively within that class. However, as patents increasingly combine subclasses from other primary classes, the class contrast approaches zero. We control for contrast when testing curvature's effects in our regression models and compare each patent's contrast and curvature, finding a positive correlation (see SI-Sec \ref{si:sec5} for details).

\textbf{Control Variables:} The patent’s application year is a set of dummy variables to distinguish the cohort of patents to estimate the invention’s value by year. The CPC section, the highest-level category that a patent's primary code belongs to, is set as dummy variables. The number of references indicates the range of subject matter a given patent covers \cite{Harhoff2003}, which is logarithm-transformed with offset 1. For the non-patent references, we create a dummy with 1 when an invention relies on them and 0 otherwise. The number of codes in a patent is also controlled because it has been studied as a predictor of patent citation impact \cite{Lerner1994}. We control for the ratio of novel pairs resulting from the adjacent possible for each patent and the logarithm-transformed combination familiarity following \cite{Fleming2001a}. Finally, we control a patent's contrast in the main analysis. When accounting for the contrast, we also include the annual count of patents assigned to the primary code, as the growth rates of citations can be correlated with the category size \cite{Lafond2019, Carnabuci2013a}. When testing the relationship with the market value and disruption, we control for forward citations within a ten-year window.

\subsection*{Analysis strategy}
Considering the characteristics of the dependent variables, we use the binary logistic regression model for the probability of a hit patent, the negative binomial regression model for the number of forward citations, and the OLS regression model for the KPSS index and the disruption index. Tables \ref{tab:tabS3} and \ref{tab:tabS4} report descriptive statistics and correlations between the main variables.

\section*{Acknowledgement}
The authors would like to acknowledge the support of the National Science Foundation Grant Award Number 2133863 and the National Research Foundation of Korea Grant funded by the Korean Government (NRF-2018S1A3A2075175). 
The authors extend our gratitude to Brian Uzzi, Hyun Ju Jung, Sameer Srivastava, Cris Moore, Douglas Erwin, Jeho Lee, and Balázs Kovács for their valuable comments. H.Y. thanks the participants of the Kellogg School of Management MORS Brownbag Seminar, Seoul National University Strategy Seminar, Berkeley Haas MORS seminar, Stanford GSB OB Seminar, MIT Sloan TIES Seminar, University of Pittsburgh at the School of Computing and Information Seminar, Computational Social Science workshop at University of Chicago, and the IU Luddy School of Informatics, Computing, and Engineering seminar for their valuable feedback during the initial stages of the manuscript. 

\bibliographystyle{naturemag} 
\bibliography{references}
\clearpage

\newpage
\section*{Supplementary Information}
\renewcommand{\thesubsection}{1.}
\subsection{Data and Sample}
\label{si:sec1}
We used two data sources for patent records from the United States Patent and Trademark Office (USPTO): the Google Patent Dataset and PatentsView. The Google Patent Dataset offers extensive historical records of assigned technological codes and their grant and application years from 1791 to 2015, including the first patent. However, it lacks detailed information, such as citations, inventor information, and market value data. Therefore, we relied on Google Patents' historical data to analyze Edison's and Tesla's patents and use PatentView for more comprehensive analyses. We analyzed 2,815,226 patents granted between 1836 and 1935, with 1084 patented by Edison and 112 by Tesla.

On the other hand, PatentsView offers more detailed information on patents granted since 1976, including details such as application and grant dates, inventor information, citations, and technology classifications. We used PatentsView and analyzed citation impacts for patents, testing our conjecture that patents with higher curvatures have an impact on the invention value. 

Figure \ref{fig:figS1} shows the process of our sample selection. Among the granted patents between 1976 and 2023, we analyzed 4,710,933 utility patents filed between 1975 and 2010. From these patents, we created control variables in term of the patent characteristics, such as patent references, grant and application years, number of codes, primary code, and contrast. Among them, 3,324,151 patents were assigned to more than one code to build a network of technology codes, which were used to construct the technology space. 

For our regression analysis, we selected 357,964 patents granted by 2013 because we used a ten-year time window for the forward citations. Using this sample, we tested the impact of curvature on the invention’s usefulness. After merging our sample and data from Kogan et al. \cite{Kogan2017}, we used 100,345 patents with the KPSS index to analyze the relationship between the patent's curvature and market value. For the analysis of the invention’s disruptiveness, we used 344,849 patents with the disruption index, which overlapped with data from \cite{FunkData}.

Patents with multiple codes are more likely to contain high-quality inventions than those with only one code when analyzed. The combinatorial process is prone to produce better outcomes than the monotonic process. We found that U.S. patents with multiple technology codes were more useful and profitable than those with only one code (see Figure \ref{fig:figS2}).
Moreover, among these patents with multiple codes, we found that patents resulting from the adjacent possibles were more useful (especially in long-term citations) and disruptive than those without them (see Figure \ref{fig:figS3}).

We drew on the CPC main-group classification, which is the second finest-grained in the technology code system. As the analysis of the finest-grained CPC classification needs extremely high costs and resources to compute a network modeling, we broke down the finest-grained codes into the second finest-grained ones. The CPC system consists of five parts from the coarsest- to finest-grained codes: section, class, subclass, main group, and subgroup. The first capital letter of the CPC codes represents a section, followed by a class symbol denoted as a two-digit number. The subclass symbol includes the section and class symbols followed by a letter. The main-group symbol is represented as a one-digit to a four-digit number; the subgroup symbol is added up to six digits. There are approximately 250,000 CPC classification entries, including the subgroup part, at the finest-grained level. 

However, a large-scale network might not be suitable for the current analysis because the operational time of a network algorithm (i.e., Dijkstra’s algorithm) increases remarkably proportional to the maximum number of links possible in a network. In other words, it is determined by the squared value of the number of nodes \cite{Dijkstra1959}. As such, we employed the second finest-grained classification, the CPC main-group codes, assuming that the properties of lower-level classifications belong to higher-level ones. For example, the CPC full code C07D 203/02 was truncated into C07D 203 at the main-group classification. Even though the CPC main-group codes are not the most detailed, they can still capture technological characteristics similar to those at the finest-grained level. The annual average number of each technology code per patent, as shown in Figure \ref{fig:figS4}, exhibited similar patterns at both the finest-grained level and the main-group level. Consequently, our analysis used 10,287 CPC main-group classification entries. 

\renewcommand{\thesubsection}{2.}
\subsection{Measures}
\label{si:sec2}

\textbf{Invention Values:}
We measured the invention’s usefulness as the probability of a hit patent, defined as being in the top 5th percentile by citations among the application year cohort. It was a dummy that took the value 1 for a hit patent. The citations were measured as the number a patent has received from subsequent patents within five- and ten-year windows after the granted year of the patent. We also used the number of forward citations within each time window for a robustness check. The underlying assumption of the measure is that patent citations are highly associated with their technological value \cite{Ahuja2001, Albert1991, Trajtenberg1990}. For the market value, we used the KPSS index constructed by Kogan et al. \cite{Kogan2017}, which measures the private value of patented inventions by estimating stock market reactions to them. 
As a proxy for the disruptiveness of an invention, we adopted the disruption index \cite{Funk2017, Wu2019a} to discern the extent to which a patented invention consolidates or destabilizes knowledge flows. The disruption index, defined as $D = p_i - p_j = (n_i -n_j)/(n_i + n_j + n_k)$, is determined by the proportion of subsequent citing patents classified into three groups: \textit{i} citing just the focal patent, \textit{j} citing both the focal patent and its references, and \textit{k} citing only the references. The disruption index is bound between -1 and 1. With an increase in this value, the patent becomes more disruptive, indicating that the focal patent disrupts existing streams and initiates new ones because future inventions can access technological predecessors only through the focal patent.

\textbf{Curvature:} 
Curvature in mathematics is defined as the degree to which a curve deviates from a straight line. In our case, the straight path is defined as the number of degrees of separation without accounting for previous co-uses. When codes have been bundled so often, we consider them close to each other, proportional to their frequent use. For example, an effective resistance between $i$ and $j$, $d_{ij} = 1/n_{ij}$ where $n_{ij}$ is the number of patents that have put together codes $i$ and $j$. 
Employing familiarity measure in \cite{Fleming2001}, effects of the distant past events on our current landscape are exponentially depreciating with their time, that is, 
\begin{equation}
d_{ij} = 1/{\sum_{\Delta_t} n_{ij}(\Delta_t) e^{-\Delta_t/t_c}}
\end{equation}
\noindent,where $n_{ij} (\Delta_t)$ counts patents that have $i$ and $j$ in $\Delta_t$ years in the application date (when analyzing Edison's and Tesla's patents, we drew on the grant date), whose impact is exponentially depreciated with characteristic time $t_c$. The longer $t_c$ is, the longer memory the system has. In our case, we followed the convention that $t_c$ is five years, meaning that the events five years ago will impact less than half of the current event's impact (basically, 1/e). Considering the typical patent citation cycle spans approximately ten years, we estimate this characteristic length as five years \cite{Fleming2001a}.

We measure how much the space of possible is curved by the massive interactions by calculating the degree of separation. 
It was measured through the following four steps. In the first step, the landscape of existing technologies was treated as a network of technology codes. Thus, we built the co-occurrence network as a $n \times n$  matrix for a given year $t$, where $n$ denotes the unique CPC codes filed from 1975 up to year $t-1$. In the second step, we weighted the network links by the inverse of the combination familiarity \cite{Fleming2001a} of each pair, measured in year $t-1$, using Eq. \ref{eq:eq1}. In the third step, we found the shortest paths of the newly recombined components at year $t$. Assuming that the components exist in the cumulative landscape, we retrieved new pairwise combinations from the networks accumulated from 1976 up to year $t-1$; the base network was one constructed by patents filed in 1975. We then observed the weighted and unweighted shortest paths of the first-ever recombined pairs, corresponding respectively to the least resistance path and the expected path without resistance, using 35 weighted networks of patented technologies by moving year $t$ from 1976 to 2010. Dijkstra's \cite{Dijkstra1959} algorithm was used to find the shortest path between the newly recombined components we wanted to analze, which is an iterative algorithm that selects a neighboring node with the minimum cost to move on to the next one in every step. At the technology pair level, the curvature was calculated by the ratio of the weighted and unweighted shortest path distances (see Eq. \ref{eq:eq2}). We defined its value as 0 when a newly recombined pair has no indirect path. Finally, our regression models included the average curvature of novel pairs used in the focal patent.

\textbf{Familiarity:} It indicates how recently and frequently existing technologies have been combined \cite{Fleming2001a}. We counted the number of prior patents that used a particular combination of technologies retrieved from a focal patent field in year $t$. This number denotes the cumulative usage of each technological combination used in the focal patent. Then, we multiplied the cumulative combination usage and an exponentially decaying component that indicates the difference between the application year of the focal patent and that of the prior patents. The latter reflects the loss of knowledge over time. The sum of the multiplied values was defined as the combination familiarity of a given technological combination. In sum, the following equation denotes combination familiarity:

\begin{equation}
\sum_{\Delta_t}  n_{ij} (\Delta_t) e^{-\Delta_t / c}
    \label{eq:familiarity}
\end{equation}

\noindent, where $n_{ij}$ ($\Delta_t$) denotes how frequently the technology codes were used $\Delta_t$ years ago. The impact of use frequency $n_{ij}$ to the familiarity measure is exponentially depreciating at the rate of $c$, the characteristic time for memory loss (or knowledge loss). Considering the typical patent citation cycle spans approximately ten years, we estimate this characteristic length as five years \cite{Fleming2001a}. We calculated its patent-level measure by taking the average value of combination familiarity for a patent. Our analysis used the logarithm of the combination familiarity average.

\textbf{Novelty Proportion:} We accounted for the ratio of the newly recombined pairs resulting from the adjacent possibilities in a given patent. It was calculated by dividing the count of these novel pairs by the total number of technology pairs in the focal patent. 

\textbf{Contrast:} Following \cite{Carnabuci2015, Kovacs2021}, we calculate the category contrast for each main-group code in the CPC classification. The contrast measure quantifies the extent to which a primary code (the first-listed CPC code on a patent) is exclusively associated with its sub-categories in various inventions. The underlying principle is that the more often subordinate codes belong to their primary code within patents, the more distinctly this primary code emerges as a representative symbol of the invention. Accordingly, a code's contrast is calculated in two steps. Initially, we compute the proportion of the main-group level codes that belong to the primary code in the subclass classification, identified as their higher-level category. Subsequently, this proportion is averaged across all patents for each year to determine the contrast level.

\textbf{Control Variables:}
We accounted for several characteristics of patent information. The patent’s application year was a set of dummy variables to distinguish the cohort of patents to estimate the invention’s value by year. The number of patent references was included to indicate the range of subject matter a given patent covers \cite{Harhoff2003}; it was logarithm-transformed with offset 1. For the non-patent references, we created a dummy with 1 when an invention relied on them and 0 otherwise. The number of CPC main-group codes in a patent was controlled because it has been studied as a predictor of patent citation impact \cite{Lerner1994}. We controlled for the novelty proportion in the main analysis. Finally, we included a patent's contrast as a control variable in the main analysis. When accounting for the contrast, we also include the annual count of patents assigned to the primary code, as the growth rates of citations can be correlated with the category size \cite{Lafond2019, Carnabuci2013a}. When testing the relationship with the market value and disruption, we controlled for forward citations within a ten-year window.

\renewcommand{\thesubsection}{3.}
\subsection{Curvatures of Edison's and Tesla’s Adjacent Possible}
\label{si:sec3}
By comparing the curvature levels of Edison’s and Tesla’s inventions, we provide more relatable examples that illustrate how new possibilities adjacent to typical assemblies can lead to varying levels of success. Thomas Edison and Nicola Tesla were competitors in their contemporaries. Edison was a well-known entrepreneur and businessman. The pursuit of his work was to solve a problem and practical use for society no matter scientific understanding \cite{Stokes1997}. While Edison’s inventing process was recursive experimentation, Tesla’s work was more based on a theoretical approach and engineering design. Tesla has been evaluated as a more original mind than Edison. On the one hand, Edison was granted 1,084 utility patents between 1869 and 1933. On the other hand, Tesla had 112 US patents at that period.

Using 2,815,226 patents with more than one USPC code, granted between 1836 and 1935, we calculated the curvature of novel pairs that had not been combined at the time. We found the first-ever recombined pairs within the networks accumulated from 1837 up to a given year; the base network was one constructed by patents granted in 1836. The network links were weighted by Eq. \ref{eq:eq1}. We then observed the shortest paths of the pairs, according to Dijkstra’s algorithm, within the accumulated networks of patented technologies by moving year t from 1837 to 1935. Subsequently, we calculated the curvature of each pair according to Eq. \ref{eq:eq2}.

We averaged the curvature of each pair by taking snapshots every ten years; for example, we labeled the decade from 1865 to 1875 as 1870. Among Edison’s patents, one hundred had curvature, whereas among Tesla’s patents, seventeen had curvature. We compared the average patent's curvature of Edison and Telsa (see Fig. \ref{fig:fig2}a) and the average curvature of each pair they used (see Fig. \ref{fig:fig2}b). Tables \ref{tab:tabS1} and \ref{tab:tabS2} reported the top five patents with the highest curvatures of Edison and Tesla. On the other hand, these are the three lowest curvatures that Edison’s patent created pairs in Fig. \ref{fig:fig2}b. (1) Patent 1234451: Mold or Transfer Plate; (2) Patent 1266780: Storage Battery; and (3) Patent 1342326: Composition of Matter for Sound-Records or the Like and Process of Making the Same. Patents 1234451 and 1342326 are related to sound recording; these innovations were too costly to encourage market growth. Patent 1266780 is for storage and portable battery, which was too expensive and significantly less efficient in low temperatures, making it less commercially valuable. In sum, we found that Edison discovered new possibilities arising from higher curvature than Tesla. These findings support the difference in Edison’s and Tesla’s exploration trajectories, which are consistent with well-known historical records in terms of their strategies and approaches to invention.

\renewcommand{\thesubsection}{4.}
\subsection{Curvatures and Values}
\label{si:sec4}

To test the curvature impacts on the invention value in regression analysis, we first assessed the possibility of multicollinearity. The generalized variance inflation factor (GVIF) was calculated because there were multi-level categorical variables in our study variables. $\text{GVIF}^{1/(2 \times df)}$ was also calculated to compare GVIFs across dimensions, where \textit{df} denotes the degree of freedom of the variable \cite{Fox1992}. As a rule of thumb for testing multicollinearity, $\text{GVIF}^{1/(2 \times df)}$ should be less than 2. This value is comparable to a variance inflation factor (VIF) threshold value of 4 as a recommended VIF value for multicollinearity \cite{Hair2018}. The $\text{GVIF}^{1/(2 \times df)}$ values of all variables in this study ranged from 1.006 to 1.437, indicating no multicollinearity issues. Tables \ref{tab:tabS3} and \ref{tab:tabS4} show the descriptive statistics and a correlation matrix for the study variables.

\subsubsection*{Curvatures and Hit Probability}
We ran several regression analyses on the effects of curvature on the invention value. Here, we used the average curvature of novel pairs in a patent to test regression models.
We found a positive correlation between the quartile of the curvature and the hit probability within a ten-year window, as shown in Figures \ref{fig:fig3}a. We ran the binary logistic regression model to estimate the hit probability because it was coded as a dummy. Table \ref{tab:tabS5} presents the regression results for predicting the variable. In Table \ref{tab:tabS5}, Model 1 presents the effects of the control variables on the hit probability. Our results in this model were consistent with what previous studies found: for example, there was a positive coefficient for the contrast ($\beta = 0.712$, $p < 0.001$). Model 2 adds the curvature. We found a positive coefficient for the curvature ($\beta = 0.135$, $p < 0.001$), indicating that a one-unit increase in the curvature is associated with a 14\% ($e^{0.135} - 1$) increase in the odds of being a hit patent within a ten-year window. Figures \ref{fig:fig3}b display the effect of curvature for predicting the hit patent probability within a ten-year window estimated by Model 2 of this table.

In the same way, Models 3 and 4 were tested for the probability of a hit patent within a five-year window, whose findings were reported as our main analysis results. Model 3 included the control variables. In Model 4, the curvature has a positive coefficient ($\beta = 0.090$, $p < 0.001$), which means that a one-unit increase in the curvature changes a 9\% ($e^{0.090} - 1$) increase in the odds of being a hit patent within a five-year window. Figure \ref{fig:figS5}a displays a positive correlation between the quartile groups and the hit probability within a five-year window. Figure \ref{fig:figS5}b displays a positive impact of curvature for predicting the hit based on the regression results in Model 4 of Table \ref{tab:tabS5}.

\subsubsection*{Curvatures and Forward Citations}
When estimating the number of forward citations, we employed the negative binomial regression model to appropriately estimate an over-dispersed parameter \cite{Hilbe2007, Long1997}. The forward citations in our patent data exhibited overdispersion with a variance greater than the mean. The forward citations were a count variable of zero or a positive number. This is consistent with previous findings that patent citations are not normally distributed \cite{Katz2016, ONeale2012}. Such a non-negative count variable violates the assumptions of linear regression models with biased coefficient estimates \cite{Long1997}. Poisson models are generally used to account for this problem. However, the Poisson distribution needs a prerequisite that the mean and variance should be equal. Following Fleming \cite{Fleming2001a}, we used the negative binomial model because the forward citations were over-dispersed and followed a negative binomial distribution.

Models 1 and 2 of Table \ref{tab:tabS6} report the results of regressions that predict the forward citations within a ten-year window. Model 1 included the control variables. Model 2 added the curvature. In Model 2, we found a positive coefficient for the study variable ($\beta = 0.062, p < 0.001$). This means that for each one-unit increase in curvature, the expected log count of forward citations within a ten-year window increases by 0.062. Models 3 and 4 of this table show the regression results on the relationship between the curvature and forward citations within a five-year window. Model 4 has a positive coefficient for the curvature ($\beta = 0.044$, $p < 0.001$). This means that for each one-unit increase in curvature, the expected log count of forward citations within a five-year window increases by 0.044.

We plotted the effects of the curvature on the forward citations. Figure \ref{fig:figS6}a shows positive correlations between the quartile group of the curvature and the forward citations within a ten-year window. Figure \ref{fig:figS6}b displays a positive relationship between the curvature and the predicted values of the forward citations within a ten-year window based on the negative binomial regression in Model 2 of Table \ref{tab:tabS6}.
We found similar findings in Figure \ref{fig:figS7} on the effects of curvature on the forward citations within a five-year window. All our results did not change with a hit probability of five years instead of ten years for the probability of a hit patent. They were also robust to the different measures, such as the total forward citations within five- and ten-year windows.

\subsubsection*{Curvatures and Market Value}
In our sample, 100,345 patents overlapped with data of Kogan et al. \cite{Kogan2017} were tested for market value. Figures \ref{fig:fig3}c and \ref{fig:fig3}d in the main manuscript show that patents with high curvatures lead to high market value (KPSS in nominal dollars), using both quartile and regression results. In Table \ref{tab:tabS7}, Model 1 included control variables and Model 2 added our main study variable. We found that, in Model 2, the curvature had a positive effect on the KPSS index in nominal dollars ($\beta = 0.021$, $p < 0.001$), which indicates that one unit increase in the curvature results in a roughly 2\% increase in the KPSS index in nominal dollars.

Our findings for market value was robust against deflated to 1982 million dollars using the Consumer Price Index (CPI) \cite{Kogan2017}. Similar to the results of the main analysis, the relationship between the curvature and the KPSS index in real dollars was positive in Model 4 of Table \ref{tab:tabS7} ($\beta = 0.020$, $p < 0.001$), meaning that one unit increase in the curvature results in a roughly 2\% increase in the KPSS index in real dollars. Figure \ref{fig:figS8}a shows a positive correlation between the quantile group of the curvature and the KPSS indices in real dollars. Figure \ref{fig:figS8}b exhibits that the curvature positively predicts the KPSS index in real dollars based on the OLS regression in Model 4 of Table \ref{tab:tabS7}. In all instances, the results consistently supported our conclusions drawn from our primary analysis.

\subsubsection*{Curvatures and Disruption}
We examined the relationship between the curvature and disruption. Merging our sample and data provided by Funk and Owen-Smith \cite{FunkData}, we used 344,849 patents to run the OLS regression models reported in Table \ref{tab:tabS8}. As the disruption index is calculated based on the flow of citations, it differs by the time window of citations. We used the disruption index within a five-year window and one calculated as of 2017, provided by Funk \cite{FunkData}. Table \ref{tab:tabS8} represents the results of regressions that predict the disruption indices. In Table \ref{tab:tabS8}, Models 1 and 2 estimate the disruption index within a five-year window. The former model included the control variables. The latter model added the explanatory variable. We found a negative coefficient for the curvature in Model 2 ($\beta = -0.002$, $p < 0.001$). In our main analysis, the insets of Figure \ref{fig:fig3}a shows negative relationships between the quartile groups of curvature and the disruption within a five-year window; that of Fig. \ref{fig:fig3}b illustrates the negative effect of curvature on the disruption.

Our findings were robust for Models 3 and 4 in terms of the disruption as of 2017. In particular, there was a negative coefficient for the disruption as of 2017 in Model 4 ($\beta = -0.001$, $p < 0.001$). Figure \ref{fig:figS9}a plots a negative correlation between the quantile group of the curvature and disruption as of 2017. Figure \ref{fig:figS9}b also shows a negative relationship between the curvature and the predicted value of the disruption based on Model 4 of Table \ref{tab:tabS8}.

\subsubsection*{Caveats and Limiations}
No empirical analysis is flawless, and it is essential to recognize its limitations. Firstly, the technology codes of patented inventions were truncated into the CPC main-group codes for our analysis. Due to the algorithm's time complexity \cite{Dijkstra1959}, we assumed that the main-group classes represent the minimum weight for identifying variations in hierarchical domains within a patented invention. While this truncated classification facilitated the mapping of the technology space, it requires cautious interpretation since the finest-grained level codes were not fully identified.

There are several potential future extensions for the technology space. It can be constructed using finer-grained classifications and their recombination processes, allowing us to examine how mechanisms differ at various observational scales. Additionally, this space could be enhanced by incorporating currently utilized Large Language Models, which would surpass human-judged classification systems \cite{Mens2023}.

Furthermore, the spatial dimension can be extended to the firms' characteristics and their R\&D capacity. Finally, our framework can be complementary to qualitative observations of individual inventive processes through survey or ethnographic analysis. Our quantitative framework offers future efforts, such as supporting and complementing qualitative research efforts to get consistent findings across different variables in terms of the invention’s practical use and market value.

\renewcommand{\thesubsection}{5.}
\subsection{Curvature and Contrast}
\label{si:sec5}
Our measurement employs a bottom-up approach to identify a domain characterized by a massive cluster shaped by the intertwining of closely associated technologies. This typically clustered assembly influences its surroundings, creating an environment conducive to successful innovation. We explain that the boundaries surrounding these typical assemblies indicate an area ripe for innovation. When a novel idea appears close to these boundaries, its technological uncertainties diminish, leveraging the well-established knowledge within the massive cluster.

Conversely, conventional approaches often arise from a top-down categorization system in the hierarchical taxonomy and classification, such as nanotechnologies under the USPC code 977. If a category includes more subordinate technologies belonging to it, it is more distinct from other categories. Such a high-contrast category offers prominent information, capturing inventors' attention and influencing invention values \cite{Kovacs2021}. This categorization also affects shaping the search for new inventive opportunities and subsequently impacts invention values.

We compared the curvature and contrast, representing bottom-up and top-down perspectives, respectively. Despite a positive correlation between these measures, as shown in Figure \ref{fig:figS10}a, it turned out that these measures showed different results in terms of value creation and capture. We categorized these measures into four groups based on quantiles, finding a positive relationship with hit probability (see Figures \ref{fig:figS10}b and \ref{fig:figS10}c) and forward citations (see \ref{fig:figS10}d and \ref{fig:figS10}e). However, Figures \ref{fig:figS10}f and \ref{fig:figS10}g show that our measure positively correlates with the market value (KPSS indices), while the contrast exhibits an inverse relationship with it. Figures \ref{fig:figS10}h and \ref{fig:figS10}i display different tendencies for the disruption index. In these figures, the curvature groups negatively correlate with the disruption within a five-year window and one as of 2017, whereas the contrast groups show an inverted U-shaped with the disruption indices.

These findings imply that inventions arising from the adjacent possibles are often viewed as conventional in the market when they possess high-contrast codes, even though they have the potential to yield substantial value in citation impact. In addition, the disruptiveness would be the most for inventions with moderate contrast codes. On the contrary, an invention with high curvature correlates with heightened profitability and practical use.

\newpage
\renewcommand{\thefigure}{S1}
\begin{figure*}[htb]
    \centering
    \includegraphics[width=\textwidth]{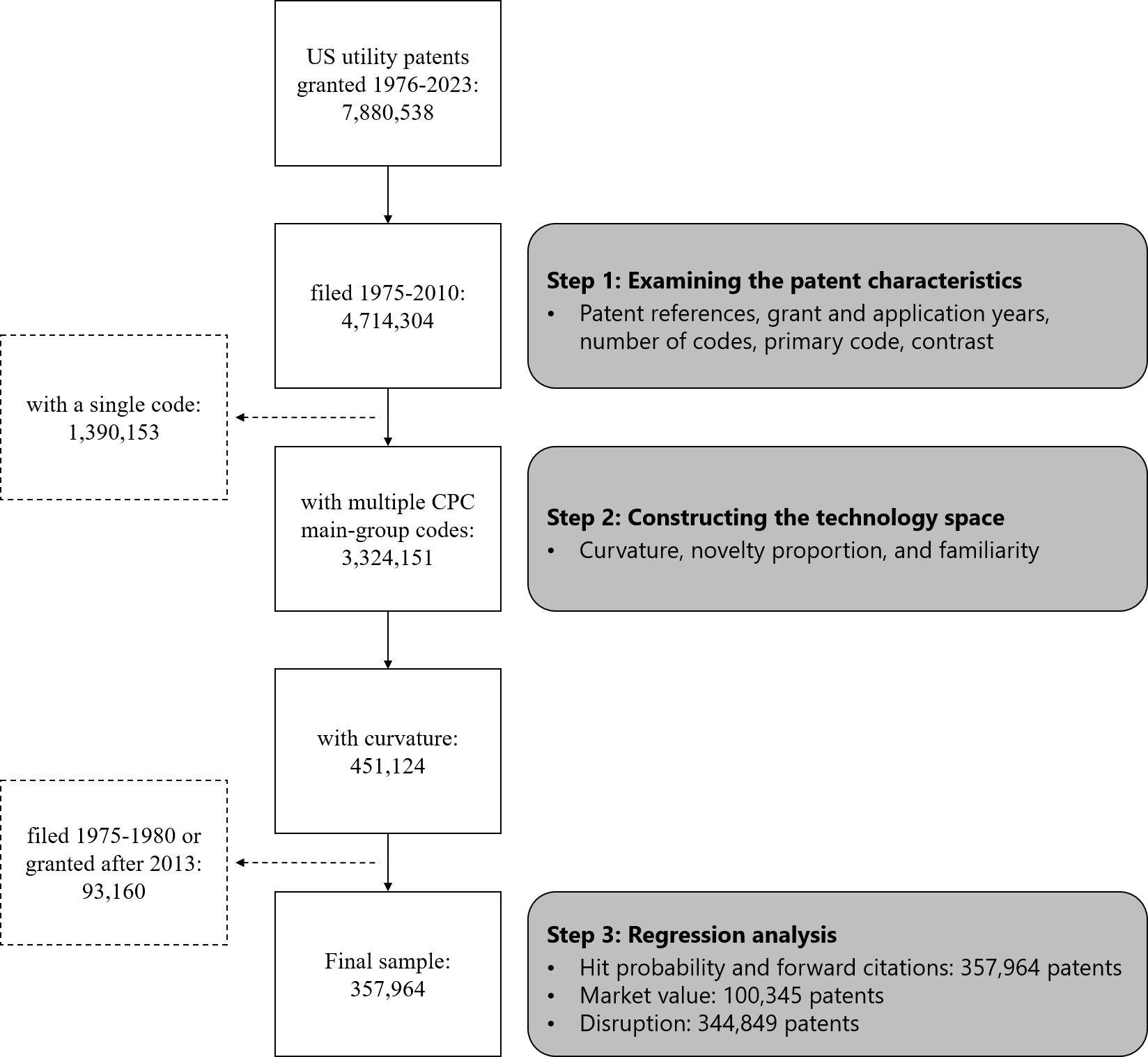}
    \caption{\textbf{Sample Selection Process.} Among 4,714,304 U.S. granted patents filed between 1975 and 2010, our regression analysis used a sample of 357,964 patents with curvature in the CPC main-group classification. 
}
    \label{fig:figS1}
\end{figure*}
\clearpage

\newpage
\renewcommand{\thefigure}{S2}
\begin{figure*}[htb]
    \centering
    \includegraphics[width=\textwidth]{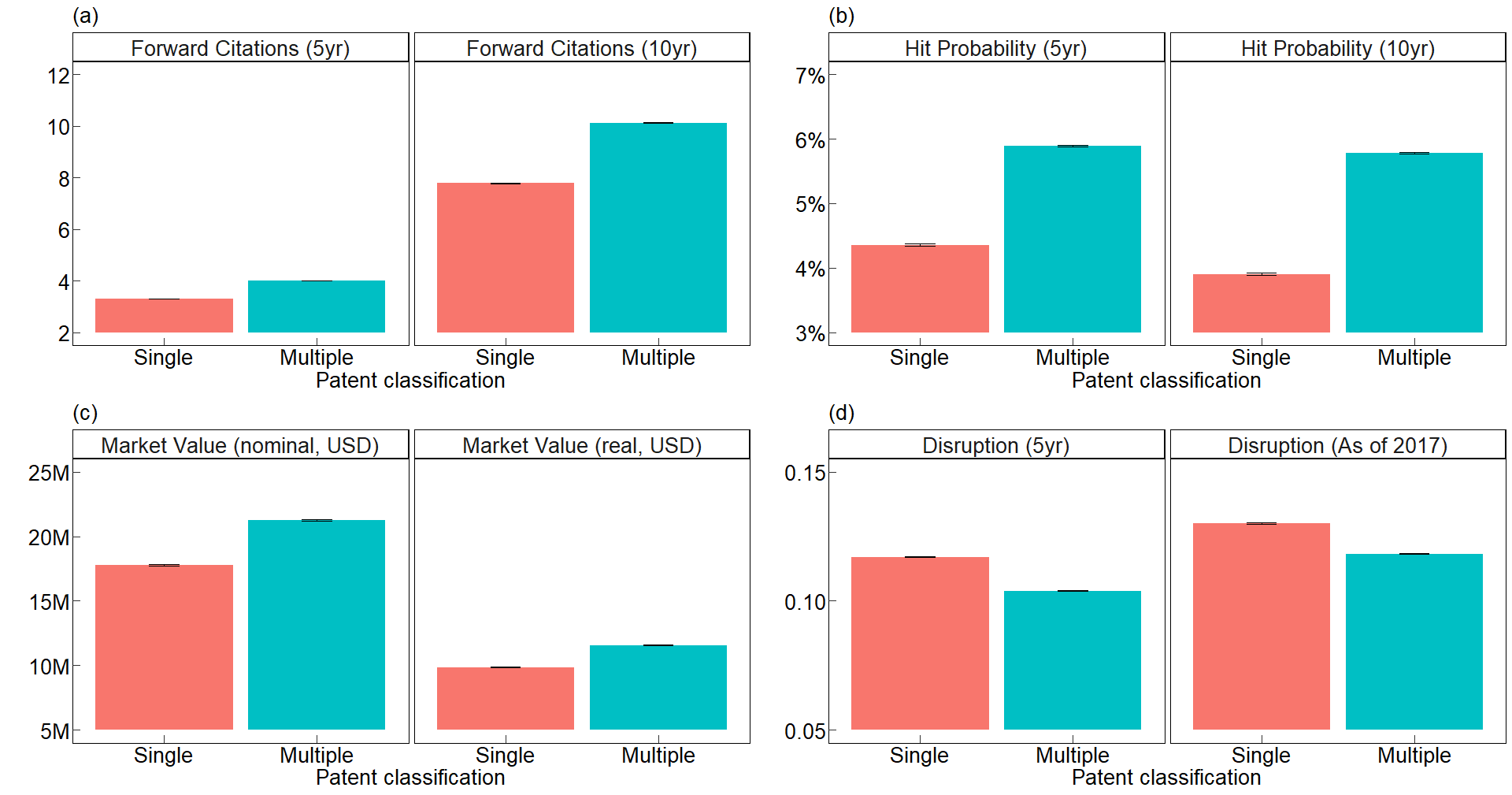}
    \caption{\textbf{Correlation for the Invention Value with Single and Multiple codes.} Patents with multiple codes received more citations than those with a single code for each time window. The results imply that an invention with a broader scope of technologies is more useful to subsequent inventions. Given a ten-year forward citation timeframe, this analysis examined 4,490,400 U.S. patents granted between 1976 and 2013 and filed between 1975 and 2010. Of these, 3,146,656 patents were assigned multiple codes at the CPC main-group level. The vertical bars represent the standard error of the mean. 
}
    \label{fig:figS2}
\end{figure*}
\clearpage

\renewcommand{\thefigure}{S3}
\begin{figure*}[htb]
    \centering
    \includegraphics[width=\textwidth]{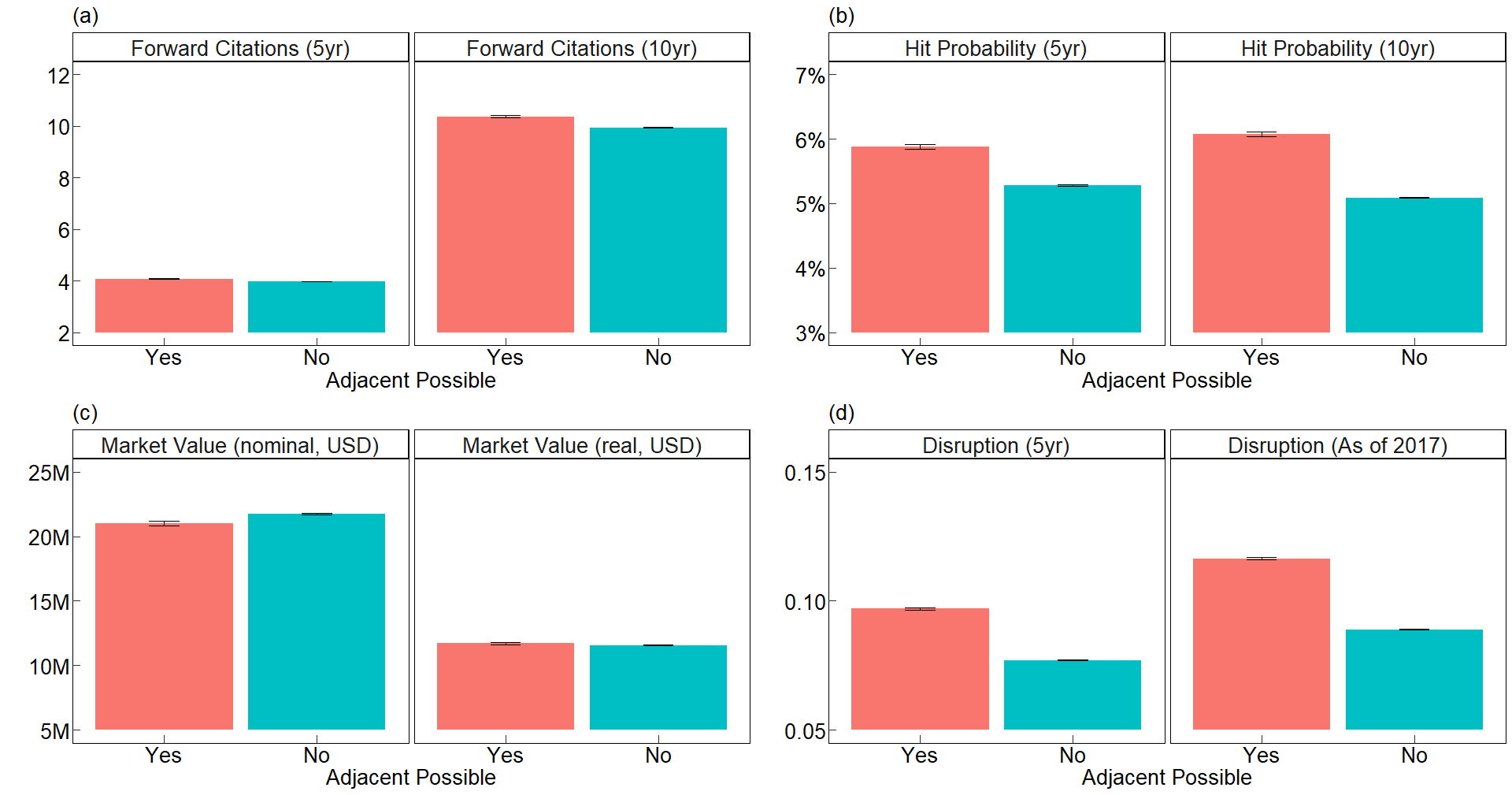}
    \caption{\textbf{Correlation for the Invention Value with the Adjacent Possibles.} \textbf{(a)} shows a positive correlation between the quantile groups of the familiarity and their average forward citations within each time window. This relationship becomes stronger for inventions with adjacent possible than those without them in long-term citations. \textbf{(b)} displays that inventions with the adjacent possibles have more chance of being a hit patent than those without it for each quantile group of the familiarity. The vertical bars represent the standard error of the mean. In this, we analyzed two groups of patents granted between 1981 and 2013 and filed between 1981 and 2010. One comprised 357,964 patents that emerged from the adjacent possible, while the other included 3,737,949 patents that did not. 
}
    \label{fig:figS3}
\end{figure*}
\clearpage

\renewcommand{\thefigure}{S4}
\begin{figure*}[htb]
    \centering
    \includegraphics[width=\textwidth]{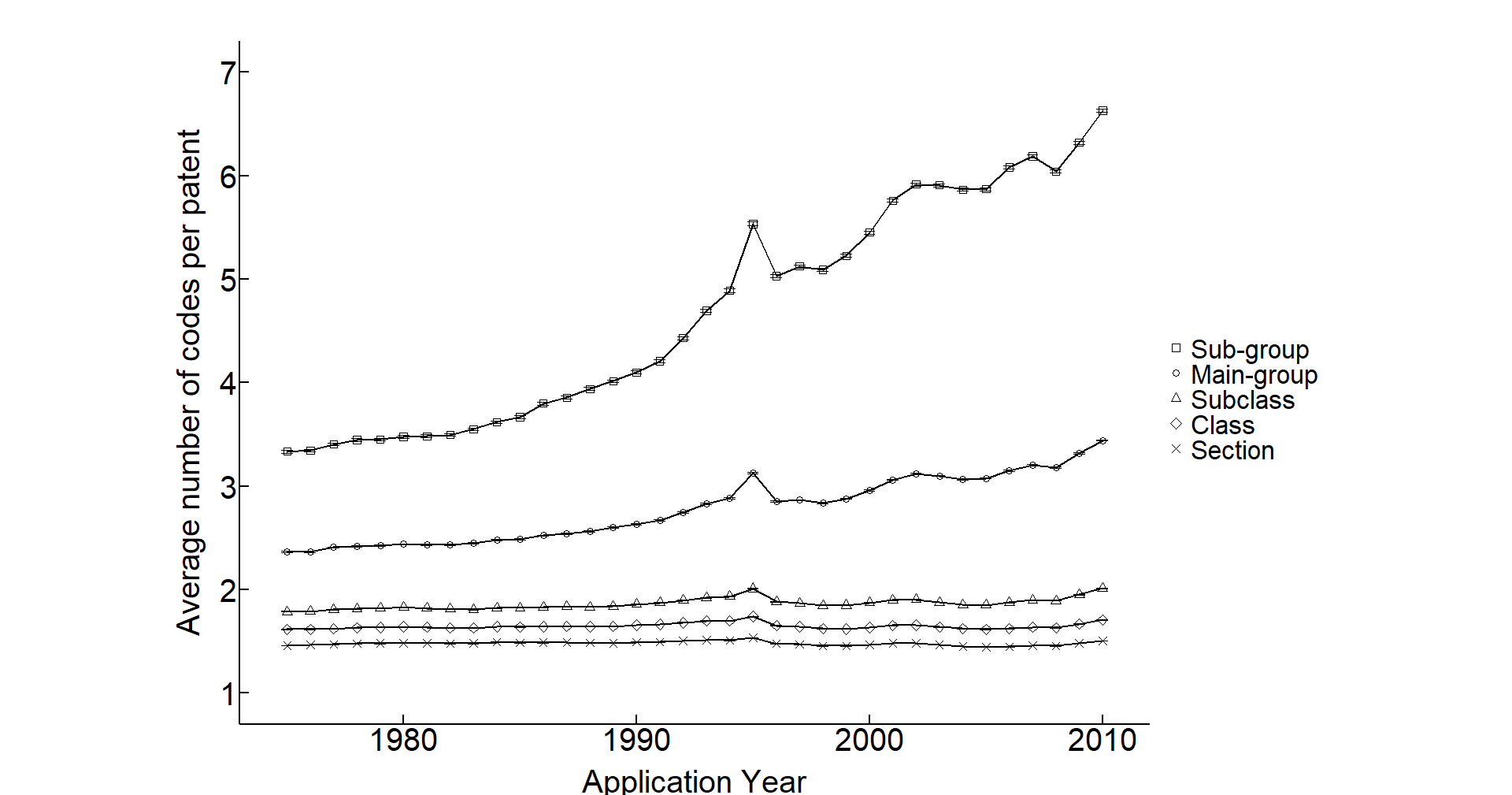}
    \caption{\textbf{Annual Average Number of CPC Codes Per Patent.} The CPC code system consists of five hierarchical parts: section, class, subclass, main group, and subgroup. While the sub-group classification as the finest-grained level has 230,277 code entries, the main-group one holds 10,287 code entries. Similar trends exist for the sub-group and main-group classifications compared to other classification levels. This study drew on the main-group classification due to the limits to computing network modeling. The plot is based on 4,710,933 US-granted patents filed between 1975 and 2010. 
}
    \label{fig:figS4}
\end{figure*}
\clearpage

\renewcommand{\thefigure}{S5}
\begin{figure*}[htb]
    \centering
    \includegraphics[width=\textwidth]{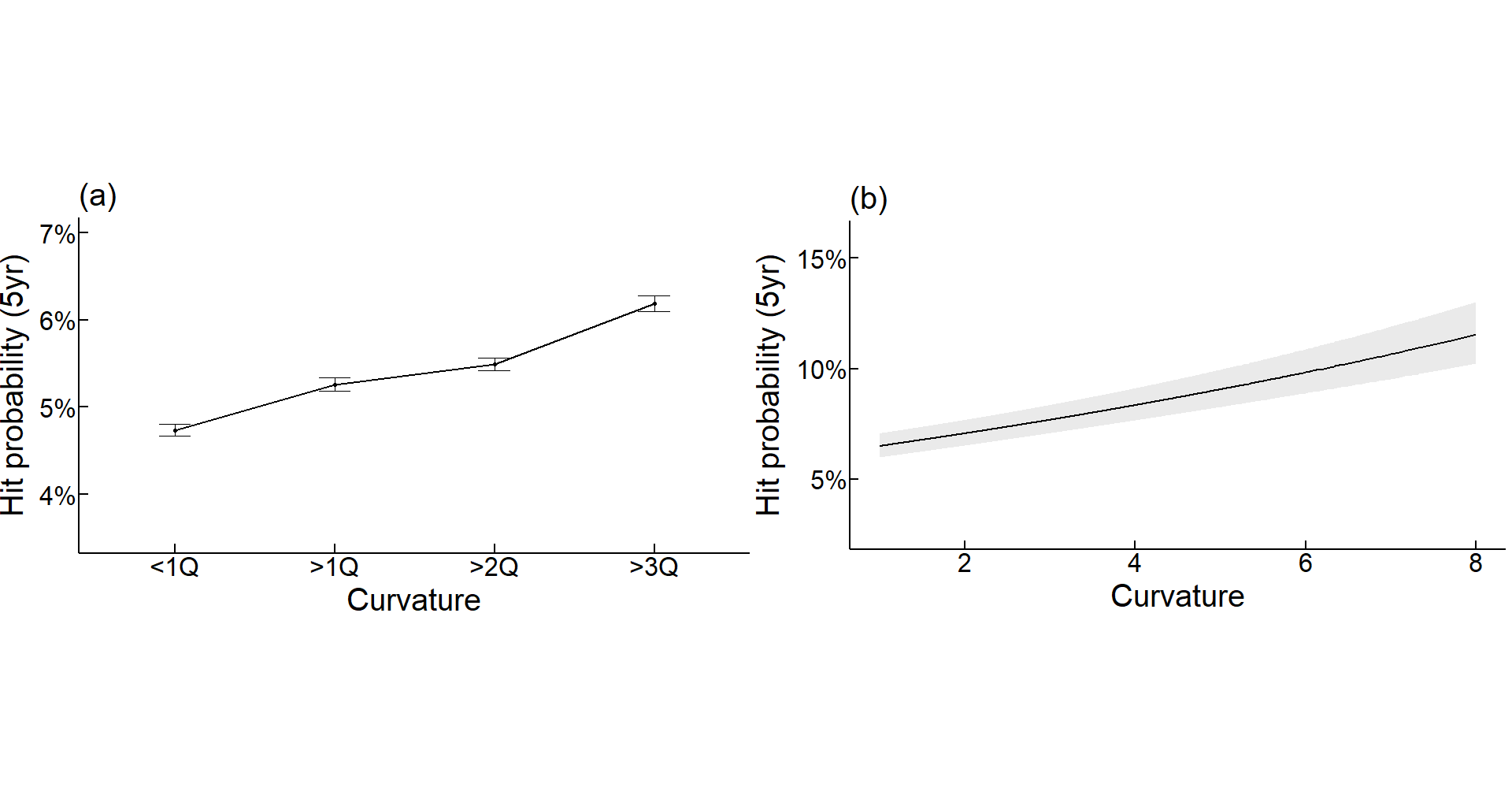}
    \caption{\textbf{Effect of the Curvature on the Hit Probability within a Five-Year Window.} \textbf{(a)} plots the correlations between the quantile group of the curvature index average and the probability of a hit patent within a five-year window. \textbf{(b)} estimates the hit probability based on the logit regression in Models 4 of Table \ref{tab:tabS5}. Vertical bars represent the mean standard error for \textbf{(a)}. Shaded areas indicate the 95\% confidence interval for \textbf{(b)}. 
}
    \label{fig:figS5}
\end{figure*}
\clearpage

\renewcommand{\thefigure}{S6}
\begin{figure*}[htb]
    \centering
    \includegraphics[width=\textwidth]{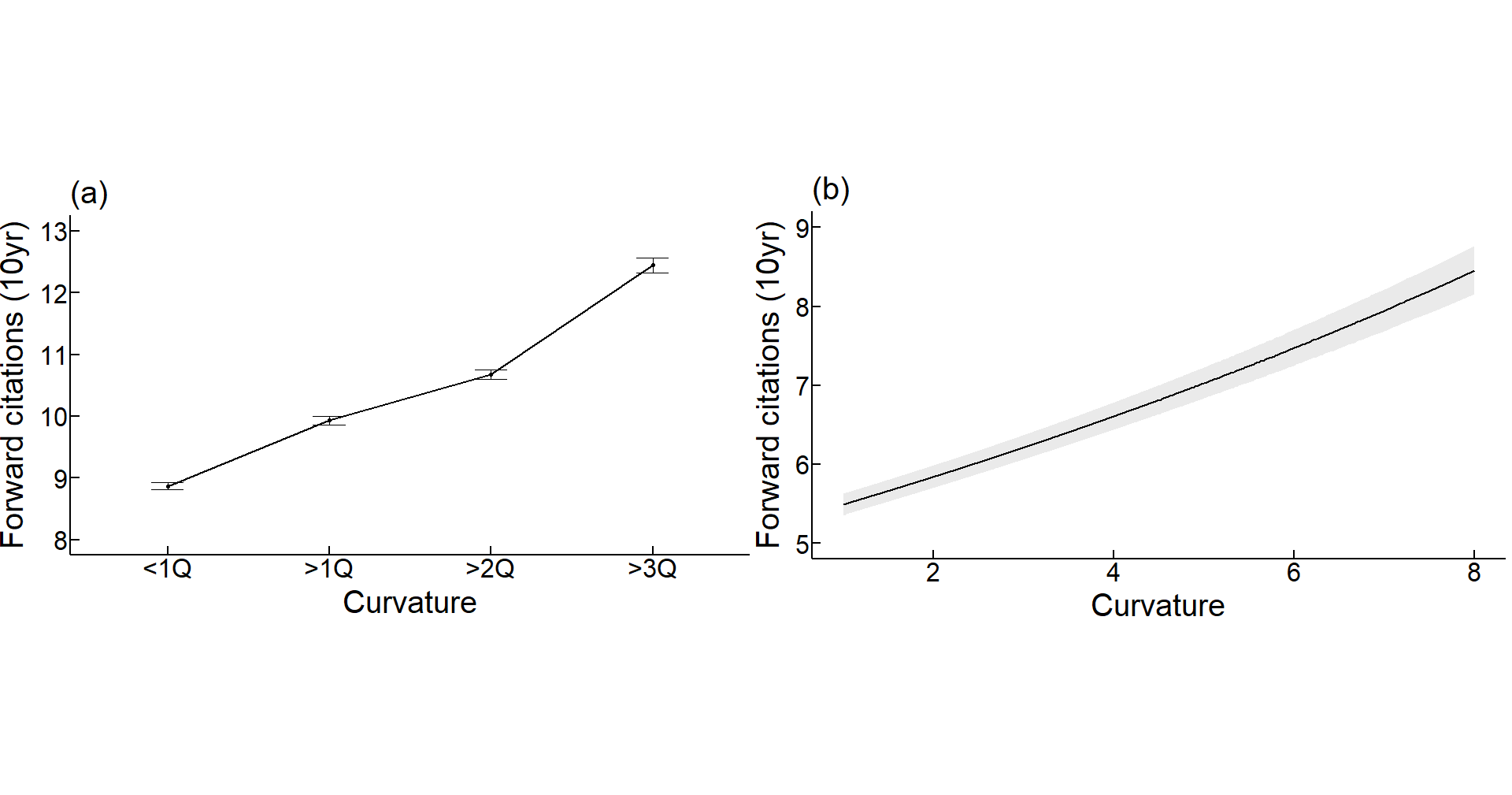}
    \caption{\textbf{Effect of the Curvature on the Forward Citations within a Ten-Year Window.} \textbf{(a)} displays the correlations between the quantile group of the curvature and the forward citations within a ten-year window. \textbf{(b)} estimates the forward citations based on the negative binomial regression in Models 2 of Table \ref{tab:tabS6}. Vertical bars represent the mean standard error for \textbf{(a)}. Shaded areas indicate the 95\% confidence interval for \textbf{(b)}.
}
    \label{fig:figS6}
\end{figure*}
\clearpage

\renewcommand{\thefigure}{S7}
\begin{figure*}[htb]
    \centering
    \includegraphics[width=\textwidth]{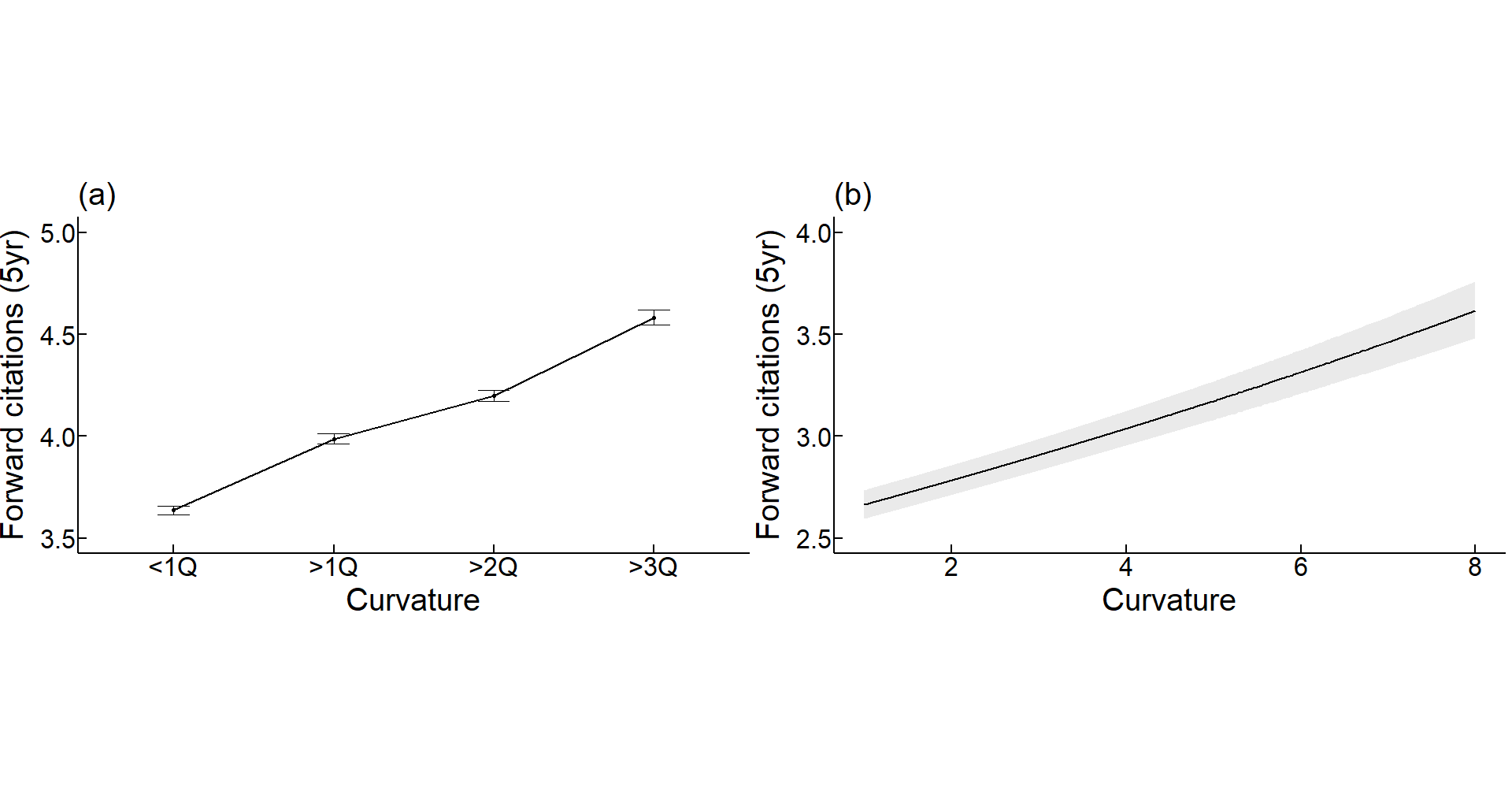}
    \caption{\textbf{Effect of the Curvature on the Forward Citations within a Five-Year Window.} \textbf{(a)} displays the correlations between the quantile group of the curvature and the forward citations within a five-year window. \textbf{(b)} estimates the forward citations based on the negative binomial regression in Models 4 of Table \ref{tab:tabS6}. Vertical bars represent the mean standard error for \textbf{(a)}. Shaded areas indicate the 95\% confidence interval for \textbf{(b)}.
}
    \label{fig:figS7}
\end{figure*}
\clearpage

\renewcommand{\thefigure}{S8}
\begin{figure*}[htb]
    \centering
    \includegraphics[width=\textwidth]{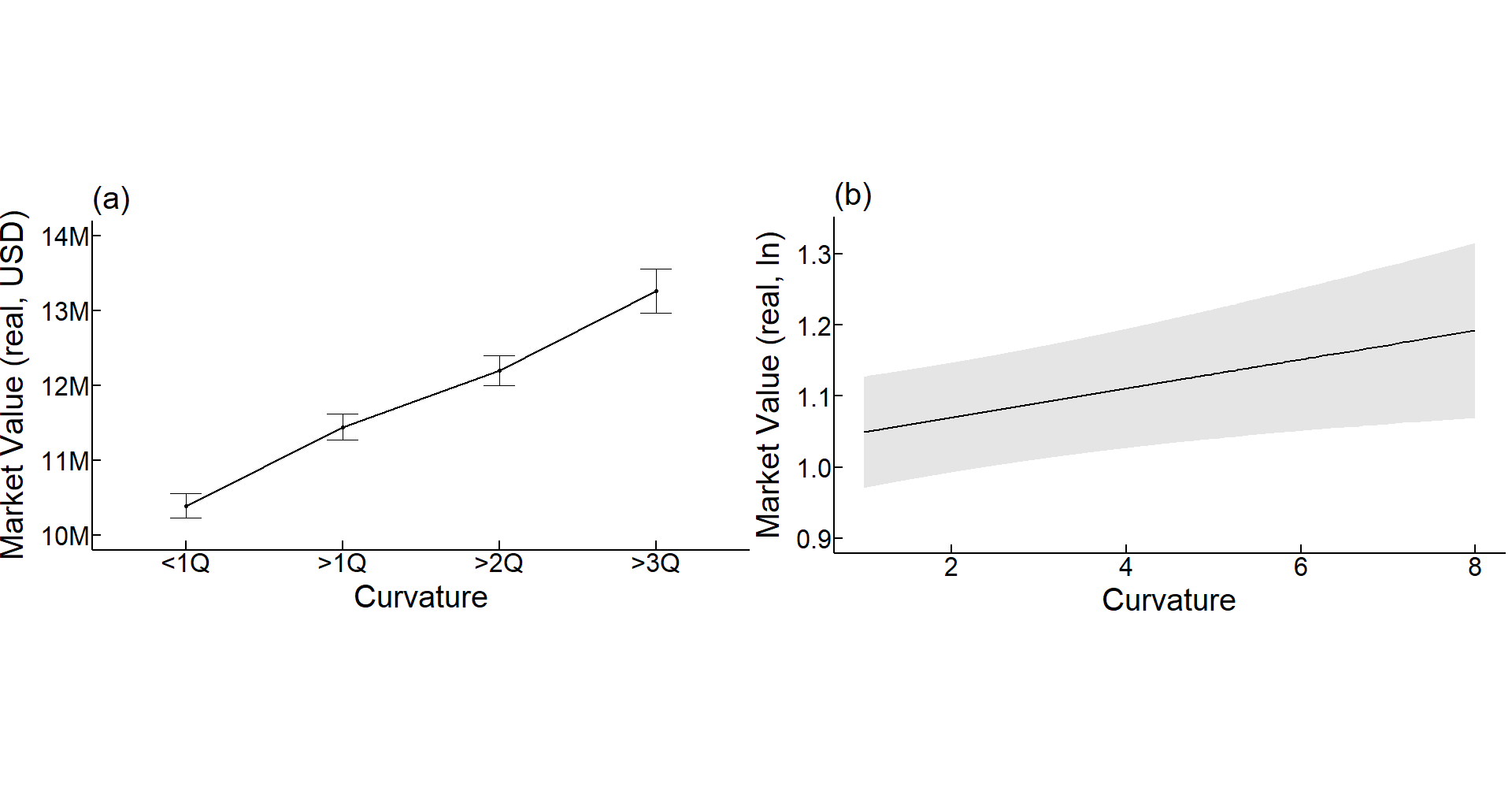}
    \caption{\textbf{Effect of the Curvature Index Average on the Market Value} \textbf{(a)} shows the correlations between the quantile group of the curvature index average and the KPSS index in real dollars. \textbf{(b)} estimates the KPSS index based on the OLS regression in Model 4 of Table \ref{tab:tabS7}. Vertical bars represent the mean standard error for \textbf{(a)}. Shaded areas indicate the 95\% confidence interval for \textbf{(b)}.
}
    \label{fig:figS8}
\end{figure*}
\clearpage

\renewcommand{\thefigure}{S9}
\begin{figure*}[htb]
    \centering
    \includegraphics[width=\textwidth]{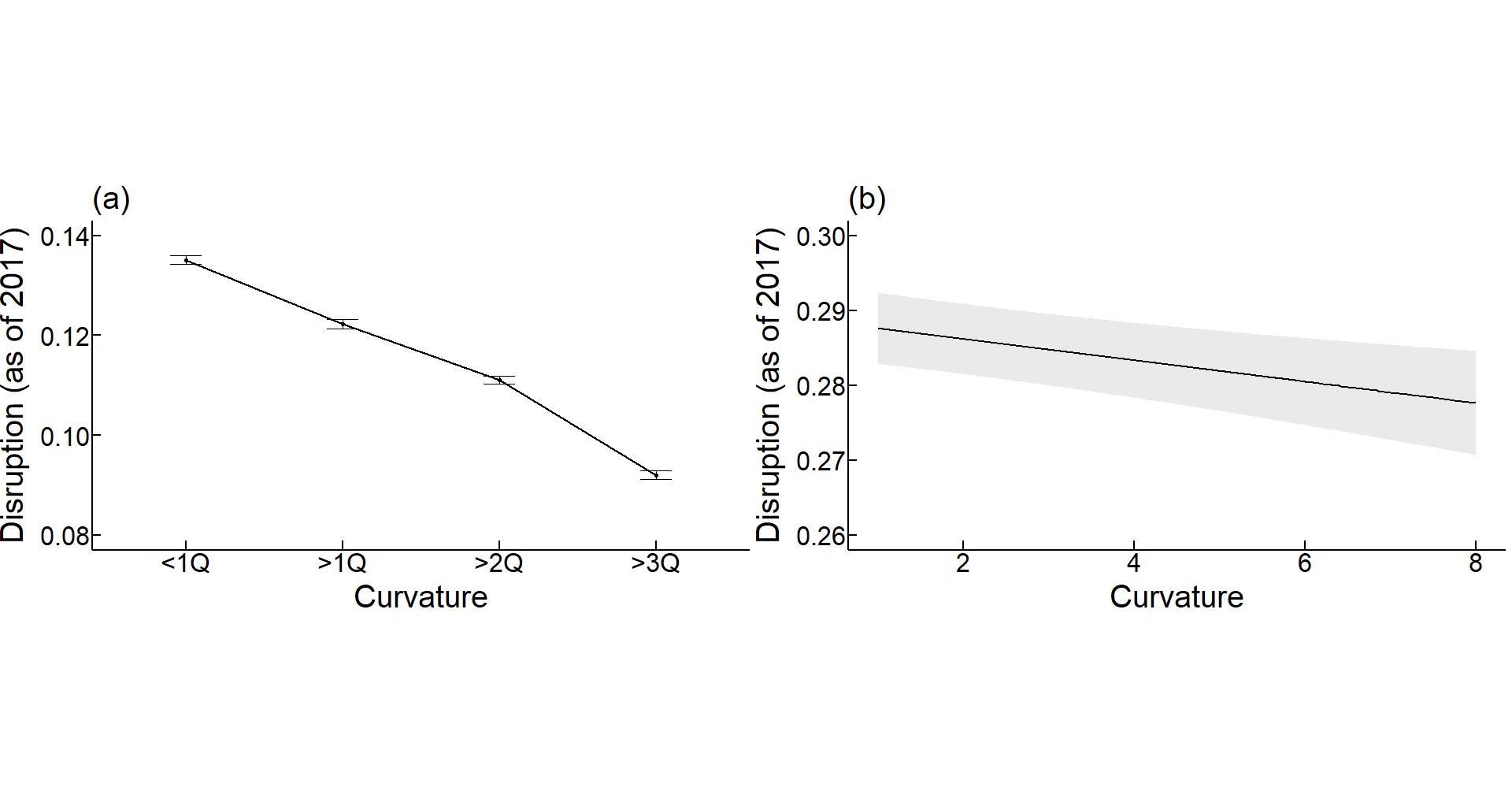}
    \caption{\textbf{Effect of the Curvature on the Disruption Index.} \textbf{(a)} shows the correlations between the quantile group of the curvature and the disruption index as of 2017. \textbf{(b)} estimates the disruption index based on the OLS regression in Models 4 of Table \ref{tab:tabS8}. Vertical bars represent the mean standard error for \textbf{(a)}. Shaded areas indicate the 95\% confidence interval for \textbf{(b)}.
}
    \label{fig:figS9}
\end{figure*}
\clearpage

\renewcommand{\thefigure}{S10}
\begin{figure*}[htb]
    \centering
    \includegraphics[width=\textwidth]{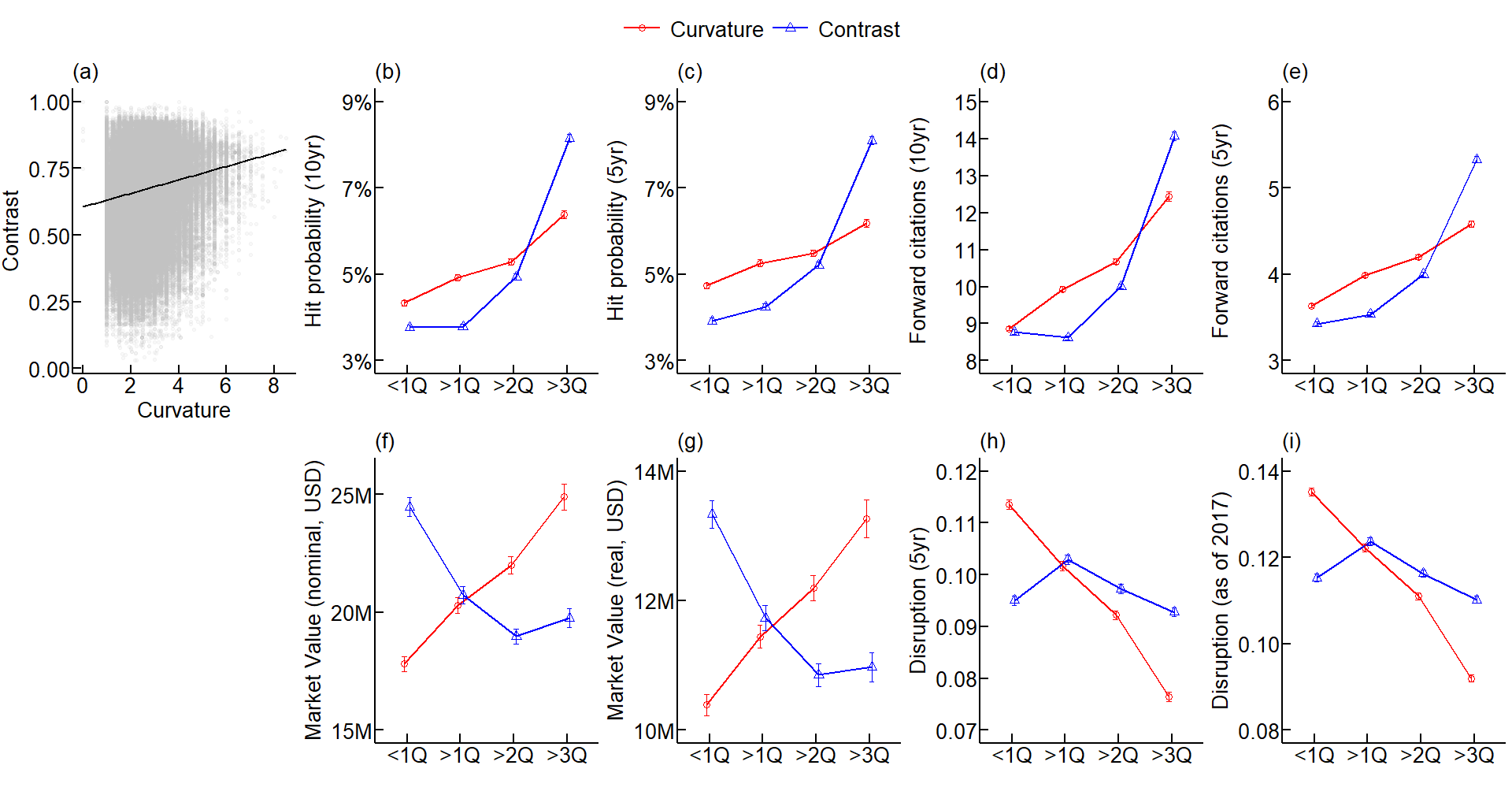}
    \caption{\textbf{Comparison between the Curvature and Contrast Measures.} \textbf{(a)} shows the correlations between the two measures. \textbf{(b)}-\textbf{(i)} illustrate the correlations between the quantile group of each measure, hit probability, forwardocations, market value, and disruption with the mean standard error. 
}
    \label{fig:figS10}
\end{figure*}
\clearpage

\newpage
\newpage
\renewcommand{\thetable}{S1}
\begin{table}[h]
\centering
\caption{Edison's Top 5 Patents with the Highest Curvature}
\label{tab:tabS1}

\resizebox{\textwidth}{!}{%

\begin{tabular}{lllll}
\hline
 & Patent ID & Grant Year & Title & Curvature \\ \hline
1 & 219628 & 1879 & Improvement in Electric Lights & 6.50 \\
2 & 251551 & 1881 & System of Electric Lighting & 5.00 \\
3 & 248565 & 1881 & Webermeter & 4.50 \\
4 & 274295 & 1883 & Incandescent Eletric Lamp & 4.25 \\
5 & 238868 & 1881 & Manufacture of Carbons for Incandescent Electric Lamps & 4.00 \\ \hline
\end{tabular}
}
\end{table}

\renewcommand{\thetable}{S2}
\begin{table}[h]
\centering
\caption{Tesla's Top 5 Patents with the Highest Curvature}
\label{tab:tabS2}
\resizebox{\textwidth}{!}{%

\begin{tabular}{lllll}
\hline
 & Patent ID & Grant Year & Title & Curvature \\
 \hline
1 & 685958 & 1901 & Method of Utilizing Radiant Energy & 3.25 \\
2 & 568177 & 1896 & Apparatus for Producing Ozone & 2.75 \\
3 & 613809 & 1898 & \begin{tabular}[c]{@{}l@{}}Method of and Apparatus for \\ \, Controlling Mechanism of Moving Vessels of Vehicles\end{tabular} & 2.50 \\
4 & 655838 & 1900 & Method of Insulating Electric Conductors & 2.50 \\
5 & 685954 & 1901 & \begin{tabular}[c]{@{}l@{}}Method of Utilizing Effects Transmitted Through \\  \, Natural Media\end{tabular} & 2.50 \\
\hline
\end{tabular}
}

\end{table}

\newpage
\renewcommand{\thetable}{S3}
\begin{table}[ht]
\centering
\caption{Descriptive Statistics}
\label{tab:tabS3}

\begin{threeparttable}
\begin{tabular}{rrrrr}
  \hline
 & mean & sd & min & max \\ 
  \hline
Hit probability (10yr) & 0.052 & 0.221 & 0.000 & 1.000 \\ 
  Hit probability (5yr) & 0.054 & 0.225 & 0.000 & 1.000 \\ 
  Forward citations (10yr) & 10.365 & 23.898 & 0.000 & 1834.000 \\ 
  Forward citations (5yr) & 4.074 & 8.126 & 0.000 & 483.000 \\
  Market Value (KPSS, nominal) \textsuperscript{a} & 20.991 & 59.784 & 0.000 & 4242.409 \\ 
  Market Value (KPSS, real) \textsuperscript{a} & 11.722 & 32.025 & 0.000 & 2379.650 \\ 
  Disruption (5yr) \textsuperscript{b} & 0.097 & 0.263 & -1.000 & 1.000 \\ 
  Disruption (as of 2017) \textsuperscript{b} & 0.116 & 0.252 & -1.000 & 1.000 \\  
  \# patent references (ln) & 2.248 & 0.908 & 0.000 & 7.659 \\ 
  Non-patent references & 0.378 & 0.485 & 0.000 & 1.000 \\ 
  \# CPC main group & 5.581 & 3.383 & 2.000 & 81.000 \\ 
  Contrast & 0.667 & 0.141 & 0.029 & 1.000 \\ 
  Primary code size (ln) & 5.884 & 1.390 & 0.000 & 9.904 \\ 
  Novelty proportion & 0.329 & 0.276 & 0.001 & 1.000 \\ 
  Familiarity (ln) & 3.268 & 1.719 & 0.000 & 9.666 \\ 
  Curvature & 2.419 & 0.921 & 0.000 & 8.500 \\ 
   \hline
\end{tabular}

\begin{tablenotes}[para, flushleft]
  Note: The number of our sampled patents was 357,964.
  
  \textsuperscript{a} $N$ = 100,345 as a subset of our sampled patents.
  
  \textsuperscript{b} $N$ = 344,849 as a subset of our sampled patents.

\end{tablenotes}

\end{threeparttable}

\end{table}

\newpage
\renewcommand{\thetable}{S4}
\begin{sidewaystable}[ht]
\centering

\caption{Correlation Matrix}
\label{tab:tabS4}

\resizebox{\textwidth}{!}{%

\begin{threeparttable}

\begin{tabular}{l
D{.}{.}{-2}
D{.}{.}{-2}
D{.}{.}{-2}
D{.}{.}{-2}
D{.}{.}{-2}
D{.}{.}{-2}
D{.}{.}{-2}
D{.}{.}{-2}
}
  \hline
 & 1 & 2 & 3 & 4 & 5 & 6 & 7 \\ 
  \hline
\# patent references (ln) &  &  &  &  &  &  &  \\ 
  Non-patent references &  0.182^{***} &  &  &  &  &  &  \\ 
  \# CPC main group &  0.144^{***} &  0.178^{***} &  &  &  &  &  \\ 
  Contrast &  0.033^{***} & -0.057^{***} & -0.221^{***} &  &  &  &  \\ 
  Primary code size (ln) &  0.050^{***} &  0.174^{***} &  0.159^{***} &  0.316^{***} &  &  &  \\ 
  Novelty proportion & -0.094^{***} & -0.133^{***} & -0.512^{***} &  0.185^{***} & -0.186^{***} &  &  \\ 
  Familiarity (ln) &  0.106^{***} &  0.187^{***} &  0.434^{***} & -0.133^{***} &  0.343^{***} & -0.644^{***} &  \\ 
  Curvature &  0.073^{***} &  0.028^{***} & -0.084^{***} &  0.165^{***} &  0.016^{***} &  0.165^{***} & -0.084^{***} \\ 
   \hline
\end{tabular}
\begin{tablenotes}[para, flushleft]
  Note: The number of our sampled patents was 357,964. The application year dummy variables were omitted.
  
  $^{\dag}$\textit{p}$<$0.10; $^{*}$\textit{p}$<$0.05; $^{**}$\textit{p}$<$0.01; $^{***}$\textit{p}$<$0.001
\end{tablenotes}

\end{threeparttable}
}

\end{sidewaystable}

\clearpage

\newpage
\renewcommand{\thetable}{S5}
\begin{table}[!htbp] \centering 
  \caption{Regression Results for Predicting the Probability of a Hit Patent} 
  \label{tab:tabS5}

\begin{threeparttable} 
\begin{tabular}{@{\extracolsep{5pt}}lD{.}{.}{-3} D{.}{.}{-3} D{.}{.}{-3} D{.}{.}{-3} } 
\\[-1.8ex]\hline 
\\[-1.8ex] & \multicolumn{2}{c}{Hit Probability (10yr)} & \multicolumn{2}{c}{Hit Probability (5yr)} \\ 
\\[-1.8ex] & \multicolumn{1}{c}{Model 1} & \multicolumn{1}{c}{Model 2} & \multicolumn{1}{c}{Model 3} & \multicolumn{1}{c}{Model 4}\\ 
\hline \\[-1.8ex] 
 Constant & -4.933^{***} & -5.171^{***} & -5.069^{***} & -5.226^{***} \\ 
   \# patent references (ln) & 0.616^{***} & 0.613^{***} & 0.631^{***} & 0.629^{***} \\ 
  Non-patent references & 0.553^{***} & 0.552^{***} & 0.528^{***} & 0.528^{***} \\ 
  \# CPC main group & 0.072^{***} & 0.071^{***} & 0.065^{***} & 0.064^{***} \\ 
  Contrast & 0.712^{***} & 0.611^{***} & 0.765^{***} & 0.696^{***} \\ 
  Primary code size (ln) & 0.099^{***} & 0.105^{***} & 0.093^{***} & 0.098^{***} \\ 
  Novelty proportion & -0.819^{***} & -0.909^{***} & -0.756^{***} & -0.814^{***} \\ 
  Familiarity (ln) & 0.044^{***} & 0.049^{***} & 0.043^{***} & 0.046^{***} \\ 
  \textbf{Curvature} &  & 0.135^{***} &  & 0.090^{***} \\ 
 \hline \\[-1.8ex] 
Application Year & Y & Y & Y & Y \\ 
Primary CPC Section & Y & Y & Y & Y \\ 
Observations & \multicolumn{1}{c}{357,964} & \multicolumn{1}{c}{357,964} & \multicolumn{1}{c}{357,964} & \multicolumn{1}{c}{357,964} \\  
McFadden's $R^{2}$ & 0.133 & 0.134 & 0.122 & 0.123 \\ 
AIC & \multicolumn{1}{c}{126,286} & \multicolumn{1}{c}{126,063} & \multicolumn{1}{c}{131,648} & \multicolumn{1}{c}{131,547} \\ 
\hline 
\end{tabular} 
\begin{tablenotes}[para, flushleft] 
  $^{\dag}$\textit{p}$<$0.10; $^{*}$\textit{p}$<$0.05; $^{**}$\textit{p}$<$0.01; $^{***}$\textit{p}$<$0.001
\end{tablenotes}

\end{threeparttable}

\end{table}

\newpage
\renewcommand{\thetable}{S6}
\begin{table}[!htbp] \centering 
  \caption{Regression Results for Predicting the Forward Citations} 
  \label{tab:tabS6}

\begin{threeparttable} 
\begin{tabular}{@{\extracolsep{5pt}}lD{.}{.}{-3} D{.}{.}{-3} D{.}{.}{-3} D{.}{.}{-3} } 
\\[-1.8ex]\hline 
\\[-1.8ex] & \multicolumn{2}{c}{Forward Citations (10yr)} & \multicolumn{2}{c}{Forward Citations (5yr)} \\ 
\\[-1.8ex] & \multicolumn{1}{c}{Model 1} & \multicolumn{1}{c}{Model 2} & \multicolumn{1}{c}{Model 3} & \multicolumn{1}{c}{Model 4}\\ 
\hline \\[-1.8ex] 
 Constant & 0.470^{***} & 0.362^{***} & -0.400^{***} & -0.477^{***} \\ 
  \# patent references (ln) & 0.341^{***} & 0.340^{***} & 0.342^{***} & 0.341^{***} \\ 
  Non-patent references & 0.224^{***} & 0.222^{***} & 0.197^{***} & 0.195^{***} \\ 
  \# CPC main group & 0.058^{***} & 0.057^{***} & 0.051^{***} & 0.050^{***} \\ 
  Contrast & 0.243^{***} & 0.201^{***} & 0.361^{***} & 0.330^{***} \\ 
  Primary code size (ln) & 0.054^{***} & 0.057^{***} & 0.048^{***} & 0.051^{***} \\ 
  Novelty proportion & -0.238^{***} & -0.273^{***} & -0.272^{***} & -0.296^{***} \\ 
  Familiarity (ln) & 0.026^{***} & 0.027^{***} & 0.021^{***} & 0.022^{***} \\ 
  \textbf{Curvature} &  & 0.062^{***} &  & 0.044^{***} \\ 
 \hline \\[-1.8ex] 
Application Year & Y & Y & Y & Y \\ 
Primary CPC Section & Y & Y & Y & Y \\ 
Observations & \multicolumn{1}{c}{357,964} & \multicolumn{1}{c}{357,964} & \multicolumn{1}{c}{357,964} & \multicolumn{1}{c}{357,964} \\ 
McFadden's $R^{2}$ & 0.038 & 0.038 & 0.039 & 0.039 \\ 
AIC & \multicolumn{1}{c}{2,292,874} & \multicolumn{1}{c}{2,292,120} & \multicolumn{1}{c}{1,707,426} & \multicolumn{1}{c}{1,707,094} \\ 
\hline 
\end{tabular} 
\begin{tablenotes}[para, flushleft] 
  $^{\dag}$\textit{p}$<$0.10; $^{*}$\textit{p}$<$0.05; $^{**}$\textit{p}$<$0.01; $^{***}$\textit{p}$<$0.001
\end{tablenotes}

\end{threeparttable}

\end{table}

\newpage
\renewcommand{\thetable}{S7}
\begin{table}[!htbp] \centering 
  \caption{Regression Results for Predicting the Market Value} 
  \label{tab:tabS7}

\begin{threeparttable} 
\begin{tabular}{@{\extracolsep{5pt}}lD{.}{.}{-3} D{.}{.}{-3} D{.}{.}{-3} D{.}{.}{-3} } 
\\[-1.8ex]\hline 
\\[-1.8ex] & \multicolumn{2}{c}{Market Value (nominal, ln)} & \multicolumn{2}{c}{Market Value (real, ln)} \\ 
\\[-1.8ex] & \multicolumn{1}{c}{Model 1} & \multicolumn{1}{c}{Model 2} & \multicolumn{1}{c}{Model 3} & \multicolumn{1}{c}{Model 4}\\ 
\hline \\[-1.8ex] 
 Constant & 0.497^{***} & 0.460^{***} & 0.466^{***} & 0.430^{***} \\ 
  Forward ciations (10yr) & 0.092^{***} & 0.092^{***} & 0.093^{***} & 0.093^{***} \\ 
  \# patent references (ln) & 0.462^{***} & 0.462^{***} & 0.457^{***} & 0.457^{***} \\ 
  Non-patent references & 0.172^{***} & 0.171^{***} & 0.162^{***} & 0.162^{***} \\ 
  \# CPC main group & 0.0005 & 0.0005 & -0.0001 & -0.0001 \\ 
  Contrast & -0.785^{***} & -0.801^{***} & -0.781^{***} & -0.796^{***} \\ 
  Primary code size (ln) & 0.050^{***} & 0.051^{***} & 0.049^{***} & 0.050^{***} \\ 
  Novelty proportion & 0.228^{***} & 0.217^{***} & 0.227^{***} & 0.215^{***} \\ 
  Familiarity (ln) & -0.037^{***} & -0.037^{***} & -0.038^{***} & -0.038^{***} \\ 
  \textbf{Curvature} &  & 0.021^{**} &  & 0.020^{*} \\ 
 \hline \\[-1.8ex] 
Application Year & Y & Y & Y & Y \\ 
Primary CPC Section & Y & Y & Y & Y \\ 
Observations & \multicolumn{1}{c}{100,345} & \multicolumn{1}{c}{100,345} & \multicolumn{1}{c}{100,345} & \multicolumn{1}{c}{100,345} \\ 
$R^{2}$ & \multicolumn{1}{c}{0.103} & \multicolumn{1}{c}{0.103} & \multicolumn{1}{c}{0.085} & \multicolumn{1}{c}{0.085} \\ 
Adjusted $R^{2}$ & \multicolumn{1}{c}{0.103} & \multicolumn{1}{c}{0.103} & \multicolumn{1}{c}{0.084} & \multicolumn{1}{c}{0.084} \\ 
\hline 
\end{tabular} 
\begin{tablenotes}[para, flushleft] 
  $^{\dag}$\textit{p}$<$0.10; $^{*}$\textit{p}$<$0.05; $^{**}$\textit{p}$<$0.01; $^{***}$\textit{p}$<$0.001
\end{tablenotes}

\end{threeparttable}

\end{table}

\newpage
\renewcommand{\thetable}{S8}
\begin{table}[!htbp] \centering 
  \caption{Regression Resutls for Predicting the Disruption} 
  \label{tab:tabS8}

\begin{threeparttable} 
\begin{tabular}{@{\extracolsep{5pt}}lD{.}{.}{-3} D{.}{.}{-3} D{.}{.}{-3} D{.}{.}{-3} } 
\\[-1.8ex]\hline 
\\[-1.8ex] & \multicolumn{2}{c}{Disruption (5yr)} & \multicolumn{2}{c}{Disruption (as of 2017)} \\ 
\\[-1.8ex] & \multicolumn{1}{c}{Model 1} & \multicolumn{1}{c}{Model 2} & \multicolumn{1}{c}{Model 3} & \multicolumn{1}{c}{Model 4}\\ 
\hline \\[-1.8ex] 
 Constant & 0.419^{***} & 0.422^{***} & 0.476^{***} & 0.478^{***} \\ 
  Forward ciations (10yr) & 0.031^{***} & 0.031^{***} & 0.031^{***} & 0.032^{***} \\ 
  \# patent references (ln) & -0.108^{***} & -0.108^{***} & -0.117^{***} & -0.117^{***} \\ 
  Non-patent references & 0.012^{***} & 0.012^{***} & 0.015^{***} & 0.015^{***} \\ 
  \# CPC main group & 0.001^{***} & 0.001^{***} & 0.002^{***} & 0.002^{***} \\ 
  Contrast & 0.001 & 0.003 & -0.004 & -0.003 \\ 
  Primary code size (ln) & 0.0002 & 0.0001 & 0.0004 & 0.0003 \\ 
  Novelty proportion & 0.020^{***} & 0.021^{***} & 0.017^{***} & 0.018^{***} \\ 
  Familiarity (ln) & -0.0003 & -0.0004 & -0.001^{**} & -0.001^{**} \\ 
  \textbf{Curvature} &  & -0.002^{***} &  & -0.001^{***} \\ 
 \hline \\[-1.8ex] 
Application Year & Y & Y & Y & Y \\ 
Primary CPC Section & Y & Y & Y & Y \\ 
Observations & \multicolumn{1}{c}{344,849} & \multicolumn{1}{c}{344,849} & \multicolumn{1}{c}{344,849} & \multicolumn{1}{c}{344,849} \\ 
$R^{2}$ & \multicolumn{1}{c}{0.183} & \multicolumn{1}{c}{0.183} & \multicolumn{1}{c}{0.245} & \multicolumn{1}{c}{0.245} \\ 
Adjusted $R^{2}$ & \multicolumn{1}{c}{0.183} & \multicolumn{1}{c}{0.183} & \multicolumn{1}{c}{0.245} & \multicolumn{1}{c}{0.245} \\ 
\hline 
\end{tabular} 
\begin{tablenotes}[para, flushleft] 
  $^{\dag}$\textit{p}$<$0.10; $^{*}$\textit{p}$<$0.05; $^{**}$\textit{p}$<$0.01; $^{***}$\textit{p}$<$0.001
\end{tablenotes}

\end{threeparttable}

\end{table}
\clearpage

\end{document}